\documentclass[%
 reprint,
 amsmath,amssymb,
 aps, onecolumn,
]{revtex4-1}

\usepackage{graphicx}
\usepackage{dcolumn}
\usepackage{bm}

\begin{document}

\preprint{APS/123-QED}

\title{The no-hair theorem and black hole shadows}

\author{Meirong Tang}%
\affiliation{%
College of Physics, Guizhou University, Guiyang 550025, China. \\
Yunnan Observatories, Chinese Academy of Sciences, 396 Yangfangwang, Guandu District, Kunming, 650216, China
}%

\author{Zhaoyi Xu}
 \email{zyxu@gzu.edu.cn}
 \affiliation{%
College of Physics, Guizhou University, Guiyang 550025, China. \\
Key Laboratory of Particle Astrophysics,
Institute of High Energy Physics, Chinese Academy of Sciences,
Beijing 100049, China.
}%

%
%

\date{\today}

\begin{abstract}
The successful observation of M87 supermassive black hole by the Black Hole Event Horizon Telescope(EHT) provides a very good opportunity to study the theory of gravity. In this work, we obtain the exact solution for the short hair black hole (BH) in the rotation situation, and calculate in detail how hairs affect the BH shadow. For the exact solution part, using the Newman-Janis algorithm, we generalize the spherically symmetric short-hair black hole metric to the rotation case (space-time lie element (2.25)). For the BH shadow part, we study two hairy BH models.
In model 1, the properties of scalar hair are determined by the parameters $\alpha_{0}$ and $L$. In model 2, the scalar hair of the BH is short hair. In this model, the shape of the BH shadow is determined by scalar charge $Q_{m}$ and $k$. 
The main results are as follows: (1) In the case of rotational short-hair BH, the value range of parameter $k$ is $k>1$ (2.25), the range of short-hair charge value $Q_{m}$ is greatly reduced due to the introduction of the BH spin $a$. When $0\leqslant Q_{m}\leqslant \frac{2}{3}\times 4^{\frac{1}{3}}$, the rotational short-hair BH has two event horizons at this time. When $Q_{m}> \frac{2}{3}\times 4^{\frac{1}{3}}$, the rotational short-hair BH has three unequal event horizons, so the space-time structure of the BH is significantly different from that of Kerr BH. (2) For model $1$, the effect of scalar hair on the BH shadows corresponds to that of $\varepsilon>0$ in reference\citep{2021JCAP...09..028K}, but the specific changes of the shadows in model $1$ are different. This is because the BH hair in reference\citep{2021JCAP...09..028K} is considered as a perturbation to the BH, while the space-time metric of model $1$ is accurate and does not have perturbation property. For model $2$, that is, the change of the BH shadow caused by short hairs, the main change trend is consistent with that of $\varepsilon<0$ in reference\citep{2021JCAP...09..028K}. Because of the special structure of the short-hair BH, the specific changes of BH shadows are different. (3) the variation of $R_{s}$ and $\delta_{s}$ with $L$ and $\alpha_{0}$ is not a monotone function in model $1$, but in model $2$, it is. These results show that scalar hairs (model $1$) have different effects on Kerr BH shadows than short hairs (model $2$), so it is possible to distinguish the types and properties of these hairs if they are detected by EHT observations. (4) as for the effects of the hairs on energy emissivity, the main results in model $1$, different energy emissivity curves have intersection phenomenon, while in model $2$ (short-hair BH), there is no similar intersection phenomenon. 
In general, various BH hairs have different effects on the shadows, such as non-monotonic properties and intersection phenomena mentioned in this work. Using these characteristics, it is possible to test the no-hair theorem in future EHT observations, so as to have a deeper understanding of the quantum effect of BHs. In future work, we will use numerical simulations to study the effects of various hairs on BHs and their observed properties. 
\end{abstract}


\maketitle


\section{Introduction}
\label{intro}

In general relativity (GR), black hole is an exact solution to Einstein's field equation, such as the common Schwarzschild black hole(e.g. see \citep{1973grav.book.....M,1998bhp..book.....F}), which describes the vacuum gravitational field near a point mass, while for spherically symmetric internal solutions, which show various forms of solutions due to different equations of state.
In the 1960s, Kerr generalized the Schwarzschild black hole to the rotational situation through precise mathematical calculation \citep{1973grav.book.....M,1963PhRvL..11..237K}. Kerr solution is extremely important to the study of black hole physics, and the physical object described by it is real in the universe. The introduction of the spin of the black hole leads to many new properties of the solution(e.g. see \citep{1973grav.book.....M,1984ucp..book.....W,1998mtbh.book.....C}). For example, Kerr black hole has an ergosphere, which makes it have a negative energy orbit, the negative energy orbits are orbits with negative angular momentum in the ergosphere of the black hole, and their rotation direction is opposite to that of the black hole; the introduction of spin parameter results in two event horizons for Kerr black hole. When the spin approaches $1$, Kerr black hole will become an extreme black hole, which is very important for the study of the quantum effect of the black hole. In classical general relativity, the solution of Einstein's field equation describes only the curvature of space-time and does not include any quantum effects.

The existence of Schwarzschild and Kerr black holes is believed to be due to the possibility of forming black holes through gravitational collapse during late stellar evolution (e.g. see \citep{1939PhRv...56..455O}) (EHT's observations of black holes will be described later). Black holes form through gravitational collapse, which has been extensively discussed in detail(e.g. see \citep{1973grav.book.....M,1977lss..conf.....H}), and physicists have found it interesting that the final result of spherically symmetric gravitational collapse is described by the mass of the system, that is, the mass of the system completely determines all the properties of spherically symmetric space-time in a vacuum. If this is generalized to Kerr black hole and Kerr-Newman black hole, then the space-time after gravitational collapse will be completely determined by mass, spin and charge of the black hole, and has nothing to do with the properties of the precursor star and the process of gravitational collapse, which is the so-called no-hair theorem\citep{1967PhRv..164.1776I,1968CMaPh...8..245I,1971PhRvL..26..331C,1972CMaPh..25..152H,1975PhRvL..34..905R,
1982JPhA...15.3173M,1995PhRvD..51.6608B,2014ApJ...784....7B,2019PhRvL.123k1102I,2021arXiv211100953W,2015PhRvL.114o1102G,2015IJMPD..2442014H}. 
The no-hair theorem is one of the main characteristics of classical black holes. Due to the extreme nature of the black hole event horizon, quantum effects near the horizon have to be considered\citep{2016PhRvL.116w1301H,2014PhRvL.112v1101H,2015CQGra..32n4001H,2013PhLB..719..419D,1992NuPhB.378..175C,1991PhST...36..258P,2017CQGra..34t4001B}. However, considering the non-trivial matter field in the black hole space-time, the black hole no-hair theorem may be violated. There are many types of black hole hairs, among which the scalar hairs are the main ones. Due to the influence of scalar hairs on black holes, the space-time metric of black holes changes. For example, J.Ovalle et al. studied black hole solutions with scalar hairs using gravitational decoupling approach\citep{2021PDU....3100744O}. In their work, scalar hairs can be reflected by mode parameters $\alpha_{0}$ and $L$. These spherically symmetric black holes with scalar hairs have been extended to the rotational case by E.Contreras et al., using gravitational decoupling method, and their basic physical properties have been calculated in detail\citep{2021PhRvD.103d4020C}. In addition, some black hole solutions with hairs have also been obtained analytically (e.g.,\citep{2011PhRvD..83l4015J,2020JCAP...09..026K,1997IJMPD...6..563D}). These approximate and exact solutions provide good conditions for physicists to understand the quantum effects of black holes.

In recent years, scientists have made a series of breakthroughs in the observation of black holes, the most important of which are the observation of gravitational waves and the measurement of the shadow of M87 supermassive black hole\citep{2016PhRvL.116f1102A,2019ApJ...875L...1E,2019ApJ...875L...5E}. In fact, the gravitational wave signal generated by the merger of black holes is enough to indicate the existence of black holes, but this is an indirect way. A more straightforward approach is to measure the black hole shadow. In April 2019, the EHT research group published the first image of the shadow of M87 black hole\citep{2019ApJ...875L...1E}, which provided the possibility for physicists to study the physical properties of the strong gravitational field. Using the observation data of EHT, scientists have conducted extensive and in-depth research on black hole physics, general relativity, etc. (e.g.,(\citep{2021PhRvD.103l2002A,2020PhRvL.125n1104P})). Recently, the accuracy of EHT measurement has been further improved, and it is very possible to further test the basic properties of black holes, such as magnetic field\citep{2021ApJ...910L..13E}.  

Some progress has been made in testing the no-hair theorem by using the observation of the black hole shadow. Mohsen Khodadi et al. studied the effect of hair on black hole shadow by using the rotational black hole with hair\citep{2021JCAP...09..028K,2015PhRvL.115u1102C}. Because the metric of the black hole with hair they used only takes into account conventional correction to Kerr black hole, the calculations, while complete, are difficult to relate the details of specific quantum hair, such as scalar hair\citep{2016PhRvL.116w1301H,2014PhRvL.112v1101H,2015CQGra..32n4001H,2013PhLB..719..419D,1992NuPhB.378..175C,1991PhST...36..258P,2017CQGra..34t4001B}. Therefore, we need to calculate the shadow shape corresponding to the black hole with the detailed scalar hairs. In reference \citep{2021JCAP...09..028K}, the authors made a few calculations on this, but did not show more details about this. In our work, we will perform detailed calculations of the shadow of the hairy black hole; using the metric of different scalar hairy black holes, we discuss how these scalar hairs change the properties of black hole shadows.  

The logical structure of this article is as follows. In section 2, we introduce the basic properties of the solutions of the black holes with scalar hairs and derive the rotational case of the metric of black holes with short hairs. In section 3, the geodesic equation of photon in rotational black hole is derived and its analytical solution is obtained. In section 4, we calculate the effect of scalar hairs on the geometric properties of black hole shadows. In section 5, we calculate how scalar hairs affect the rate of energy emission. The sixth section is the summary and discussion of the whole paper.

\section{No-hair theorem and black hole space-time }
\label{2}

\subsection{Model 1: Hairy black hole}

In classical general relativity, black holes carry no charge other than mass, charge and spin, which is known in black hole physics as no-hair theorem\citep{1967PhRv..164.1776I,1968CMaPh...8..245I,1971PhRvL..26..331C,1972CMaPh..25..152H,1975PhRvL..34..905R,
1982JPhA...15.3173M,1995PhRvD..51.6608B,2014ApJ...784....7B,2019PhRvL.123k1102I,2021arXiv211100953W,2015PhRvL.114o1102G,2015IJMPD..2442014H}. However, it is possible that the interaction of black hole space-time with matter introduces other charges, such as internal norm symmetries and certain fields, which could make it possible for black holes to carry hairs. When these corresponding physical effects are introduced into the space-time of black hole, i.e., to make black hole hairy, the presence of these hairs will change the space-time background of black hole, thus bringing the quantum effects of black hole into the classical space-time geometry. Using extended gravitational decoupling (EGD) approach, J.Ovalle et al. obtained a spherically symmetric space-time metric with hair\citep{2021PDU....3100744O}, because in the EGD approach, there is no certain matter field, which makes the hairy black holes obtained by them have great generality. In EGD, the corresponding Einstein field equation can be expressed as follows:  

\begin{equation}
R_{\mu\nu}-\dfrac{1}{2}R g_{\mu\nu}=8\pi\tilde{T}_{\mu\nu},
\label{1}
\end{equation}

where $R_{\mu\nu}$ is Ricci curvature tensor, $R$ is Ricci curvature scalar, $g_{\mu\nu}$ is the space-time metric, $\tilde{T}_{\mu\nu}$ is the total energy-momentum tensor, and can be written as

\begin{equation}
\tilde{T}_{\mu\nu}=T_{\mu\nu}+\theta_{\mu\nu}
\label{2}
\end{equation}

where $T_{\mu\nu}$ is the energy-momentum tensor in GR, $\theta_{\mu\nu}$ is the energy-momentum tensor caused by a quantum field or gravitational branch. According to general relativity, $G_{\mu\nu}=R_{\mu\nu}-\frac{1}{2}Rg_{\mu\nu}$ needs to satisfy the Bianchi identity, namely $\bigtriangledown_{\mu}\tilde{T}^{\mu\nu}=0$. When $\theta_{\mu\nu}=0$, i.e., without considering hairs, it can be proved that the solution of field equation (\ref{1}) degenerates into Schwarzschild black hole solution. By proper treatment of the tensor $\theta_{\mu\nu}$, the spherically symmetric solution of Einstein field equation can be obtained \citep{2021PDU....3100744O}  

\begin{equation}
ds^{2}=-f(r)dt^{2}+\dfrac{1}{g(r)}dr^{2}+r^{2}d\Omega^{2}
\label{3}
\end{equation}

where $d\Omega^{2}=d\theta^{2}+\sin^{2}\theta d\phi^{2}$, $f(r)$ and $g(r)$ are the metric coefficients, and the expressions in this case are

\begin{equation}
f(r)=g(r)=1-\dfrac{2M}{r}+\alpha_{0} \exp\left({-\frac{r}{M-\frac{\alpha_{0}L}{2}}}\right)
\label{4}
\end{equation}

Here, $\alpha_{0}$ describes the deformation parameters due to the consideration of hairs, so the size of $\alpha_{0}$ represents the physical meaning related to the strength of hairs. $L$ is a constant with the dimension of distance. In this work, we limit the values of $\alpha_{0}$ and $L$ as follows \citep{2021PDU....3100744O, 2021PhRvD.103d4020C}:  

\begin{equation}
0 \leq \alpha_{0} \leq 1
\label{5}
\end{equation}

\begin{equation}
0 \leq L \leq 2
\label{6}
\end{equation}

By adjusting the parameters $\alpha_{0}$ and $L$, we can analyse how hairs change the properties of the space-time metric of black holes. Meanwhile, the space-time metric provides a basis for testing the no-hair theorem of black holes from an astronomical perspective. 

Using gravitational decoupling approach, E.Contreras et al. generalized the spherically symmetric black hole solution to the rotational case\citep{2021PhRvD.103d4020C}. By applying some restrictions on $S_{\mu\nu}$ in the energy-momentum tensor $\tilde{T}_{\mu\nu}=T_{\mu\nu}+S_{\mu\nu}$, such as meeting the strong energy condition (SEC), the space-time metric of the black hole with hair was obtained  

\begin{equation}
ds^{2}=-\left[1-\dfrac{2rm(r)}{\rho^{2}}\right]dt^{2}-\dfrac{4arm(r)\sin^{2}\theta}{\rho^{2}}dtd\phi+\dfrac{\rho^{2}}{\Delta}dr^{2}+\rho^{2}d\theta^{2}+\dfrac{\Sigma \sin^{2}\theta}{\rho^{2}}d\phi^{2}
\label{7}
\end{equation}
where, the expressions of related symbols are

\begin{equation}
\rho^{2}=r^{2}+a^{2}\cos^{2}\theta
\label{8}
\end{equation}

\begin{equation}
\Delta=r^{2}-2r\tilde{m}(r)+a^{2}=r^{2}\left[ 1+\alpha_{0} \exp\left({-\frac{r}{M-\frac{\alpha_{0}L}{2}}}\right) \right]-2Mr+a^{2}
\label{9}
\end{equation}

\begin{equation}
\Sigma=(r^{2}+a^{2})^{2}-a^{2}\Delta \sin^{2}\theta
\label{10}
\end{equation}

\begin{equation}
f(r)=g(r)=1-\dfrac{2\tilde{m}(r)}{r}
\label{11}
\end{equation}

In these expressions, $a$ is the spin of the black hole, and other parameters have the same physical meaning as in the spherically symmetric case. Since rotational black holes are real in the universe, it is more feasible to test the no-hair theorem of black holes using observational data such as EHT and gravitational waves.  

In addition, to generalize the rotational hairy black hole to the charged case\citep{2021PDU....3100744O}, we only need to replace the metric coefficient in the spherically symmetric case by  

\begin{equation}
f(r)=g(r)=1-\dfrac{2M}{r}+\dfrac{Q^{2}}{r^{2}}-\dfrac{\alpha_{0}}{r} \left( M-\dfrac{\alpha_{0}L}{2} \right) \exp\left({-\frac{r}{M-\frac{\alpha_{0}L}{2}}}\right)
\label{12}
\end{equation}

where $Q$ is the charge carried by the black hole. The form of Kerr-Newman black hole with hairs is similar to (\ref{7}) $\sim$ (\ref{11}).

\subsection{Model 2: Short hair black hole}

When considering the coupling of gravity with some anisotropic fluids in general relativity, the black hole solution with hair can be obtained. These anisotropic fluids satisfy the following conditions, that is, in some conditions, the gravity will produce de Sitter and Reissner-Nordstron (RN) black holes, and in other conditions, the solution of the black hole with hair can be obtained\citep{1997IJMPD...6..563D}, whose linear element expression is  

\begin{equation}
f(r)=g(r)=1-\dfrac{2M}{r}+\dfrac{Q_{m}^{2k}}{r^{2k}},
\label{13}
\end{equation}

$Q_{m}$ is the strength parameter of the hair. When $k=1$, the space-time metric degrades into an RN black hole; when $k>1$, the space-time metric is a short-hair black hole, and $Q_{m}$ is the charged value of the short hair. In this work, we discuss the case of $k=\frac{3}{2}$ in order to understand the properties of short-hair black hole. For the black hole corresponding to the metric (\ref{13}), its energy density and pressure are respectively  

\begin{equation}
\rho=\dfrac{Q_{m}^{2k}(2k-1)}{8\pi r^{2k+2}},
\label{14}
\end{equation}

\begin{equation}
p=\dfrac{Q_{m}^{2k}(2k-1)k}{8\pi r^{2k+2}}.
\label{15}
\end{equation}

As discussed in the original literature, the black hole solution satisfies three classical conditions, namely the weak energy condition (WEC), the energy density decays faster than $r^{-4}$, and $T=T_{ab}g^{ab}\leqslant 0$.  Therefore, this black hole does not violate the no-short-hair theorem.

Next, based on the metric of spherically symmetric short-hair black hole, we derive the solution of rotating short-hair black hole using Newman-Janis (NJ) method. In short, the NJ method is to generalize spherically symmetric space-time to rotational space-time by complex transformation. The key is to simplify Einstein field equation into a second-order partial differential equations. If there is an analytic solution to this equations, the rotation form of the corresponding space-time metric can be obtained. The content of the NJ method can be referred to the relevant literature (e.g.,\citep{Newman:1965tw,2014PhRvD..90f4041A,2014PhLB..730...95A}). Follow the idea of this method, we will show the key derivation.

Since the Schwarzschild coordinate system is used in the space-time form (\ref{3}), if the space-time considered is a black hole, there is coordinate singularity in this kind of coordinate system. Therefore, in the NJ algorithm, this coordinate system needs to be transformed to the advanced null coordinates (ANC) $(u,r,\theta, \varphi)$, and the coordinate transformation is      

\begin{equation}
du=dt-\dfrac{dr}{f(r)g(r)}=dt-\dfrac{dr}{\left(1-\dfrac{2M}{r}+\dfrac{Q_{m}^{2k}}{r^{2k}}\right)^{2}},
\label{16}
\end{equation}

In ANC, we can choose the advanced null basis vector $(e^{\mu},n^{\mu},m^{\mu}, \bar{m}^{\mu})$ to expand the space-time metric, that is, the inverse form of the space-time metric is expressed as a linear combination of the basis vector, namely $g^{\mu\nu}=-e^{\mu}n^{\nu}-e^{\nu}n^{\mu}+m ^{\mu}\bar{m}^{\nu}+m^{\nu}\bar{m}^{\mu}$. For the metric (\ref{3}) considered here, its basis vector in ANC is  

\begin{equation}
\begin{aligned}
& L^{\mu}=\delta_{r}^{\mu},    \\
& n^{\mu}=\delta_{\mu}^{\mu}-\dfrac{1}{2}\left(1-\dfrac{2M}{r}+\dfrac{Q_{m}^{2k}}{r^{2k}}\right)^{2},   \\
& m^{\mu}=\dfrac{1}{\sqrt{2}r}\delta_{\theta}^{\mu}+\dfrac{i}{\sqrt{2}r\sin\theta}\delta_{\phi}^{\mu},   ~~~
 \bar{m}^{\mu}=\dfrac{1}{\sqrt{2}r}\delta_{\theta}^{\mu}-\dfrac{i}{\sqrt{2}r\sin\theta}\delta_{\phi}^{\mu}. 
\end{aligned}
\label{17}
\end{equation}

To generalize spherically symmetric space-time to the rotation case, we can perform the following operations in ANC. In complex space, the coordinate $(u,r)$ is rotated by an angle of $\theta$, that is, $u\rightarrow u-ia\cos\theta, r\rightarrow r+ ia\cos\theta$; where $a$ is a constant, which can also be interpreted as the spin of the space-time, and $\theta$ is the rotation angle. At this time, the spherical symmetry metric coefficients $f(r)\rightarrow F(r,\theta,a), g(r)\rightarrow G(r,\theta,a)$ and $h(r)\rightarrow \psi(r,\theta,a)$. $h(r)$ is the coefficient of $d\Omega^{2}$ in the metric (\ref{3}), and here $h(r)=r^{2}$. Through these operations, the basis vectors in ANC become functions expressed by $F(r,\theta,a)$, $G(r,\theta,a)$ and $\psi(r,\theta,a)$. After calculation, it is found that the components of the inverse metric $g^{\mu\nu}$ are as follows:  

\begin{equation}
\begin{aligned}
& g^{uu}=\dfrac{a^{2}\sin^{2}\theta}{\psi}, ~~~ g^{\theta\theta}=\dfrac{1}{\psi}, ~~~ g^{ur}=g^{ru}=\sqrt{\frac{G}{F}}-\dfrac{a^{2}\sin\theta}{\psi}, \\
& g^{\phi\psi}=\dfrac{1}{\psi\sin^{2}\theta}, ~~~ g^{u\phi}=g^{\phi u}=\dfrac{a}{\psi}, ~~~ g^{r\phi}=g^{\phi r}=\dfrac{a}{\psi}, ~~~ g^{rr}=G+\dfrac{a\sin^{2}\theta}{\psi}.
\end{aligned}
\label{18}
\end{equation}

By processing these inverse metric, we can get the covariance metric tensor, and thus obtain the expression of its line element as 

\begin{equation}
ds^{2}=-Fdu^{2}+2\sqrt{\frac{F}{G}}dudr+2a\sin^{2}\theta \left(\sqrt{\frac{F}{G}}+F\right)dud\phi-2a\sin^{2}\theta \sqrt{\frac{F}{G}}drd\phi  $$$$
+\psi d\theta^{2}-\sin^{2}\theta \left(-\psi+a^{2}\sin^{2}\theta \left(2\sqrt{\frac{F}{G}}+F\right)\right)d\phi^{2}
\label{19}
\end{equation}

Next, using coordinate transformation, the space-time metric can be transformed from the Eddington-Finkelstin coordinates (EFC) to the Boyer-Lindquist coordinate (BLC)(???), and the final result is   

\begin{equation}
ds^{2}=-\dfrac{\psi}{\rho^{2}}\left(1-\dfrac{2\bar{f}}{\rho^{2}} \right)dt^{2}+\dfrac{\psi}{\Delta}dr^{2}-\dfrac{4a\bar{f}\psi\sin^{2}\theta}{\rho^{4}}dtd\phi+\psi d\theta^{2}+\dfrac{\psi\Sigma\sin^{2}\theta}{\rho^{4}}d\phi^{2}
\label{20}
\end{equation}

Here, the relationship of the related symbols is  

\begin{equation}
\begin{aligned}
& k(r)=h(r)\sqrt{\dfrac{f(r)}{g(r)}}=r^{2}, ~~~ \rho^{2}=k(r)+a^{2}\cos^{2}\theta,   \\
& \bar{f}=\dfrac{1}{2}k(r)-\dfrac{1}{2}h(r)f(r), \\
& \Delta(r)=r^{2}f(r)+a^{2}, ~~~ \Sigma=\left(k(r)+a^{2}\right)^{2}-a^{2}\Delta(r)\sin^{2}\theta.
\end{aligned}
\label{21}
\end{equation}

In this way, we get the general form of the metric of the rotational space-time in the NJ method. However, there is an unknown function $\psi$ in the metric expression, which needs to be solved by Einstein field equation. Since the space-time metric at this time satisfies rotation symmetry, the component of the Einstein tensor $G_{r\theta}=0$. Meanwhile, the metric (\ref{20}) should also satisfy Einstein field equation\citep{2014PhRvD..90f4041A,2014PhLB..730...95A}. Through a series of calculations, the equation of the gravitational field can be simplified into the following equations 

\begin{equation}
\left(k(r)+a^{2}y^{2}\right)^{2}\left(3\dfrac{\partial\psi}{\partial r}\dfrac{\partial\psi}{\partial y^{2}}-2\psi\dfrac{\partial^{2}\psi}{\partial r \partial y^{2}} \right)=3a^{2}\dfrac{\partial k}{\partial r}\psi^{2}
\label{22}
\end{equation}

\begin{equation}
\psi \left( \left( \dfrac{\partial k}{\partial r} \right)^{2}+k \left(2-\dfrac{\partial^{2}k}{\partial r^{2}}\right)-a^{2}y^{2}\left(2+\dfrac{\partial^{2}k}{\partial r^{2}}\right)\right)+\left(k+a^{2}y^{2}\right)\left(4y^{2}\dfrac{\partial\psi}{\partial y^{2}}-\dfrac{\partial k}{\partial r}\dfrac{\partial \psi}{\partial r} \right)=0
\label{23}
\end{equation}

For the short-hair black hole considered in this section, $f(r)=g(r)=1-\frac{2M}{r}+\frac{Q_{m}^{2k}}{r^{2k}}, h(r)=r^{2}$, we can obtain $k(r)=h(r)\sqrt{\frac{f(r)}{g(r)}}=r^{2}$. Substituting $k(r)$ into equations (\ref{22}) and (\ref{23}), by solving the equations, we can get 

\begin{equation}
\psi(r,\theta,a)=r^{2}+a^{2}\cos^{2}\theta,
\label{24}
\end{equation}

So far, we have the expressions for all the unknown functions, thereby obtaining the rotation form of the short-hair black hole, which is 

\begin{equation}
ds^{2}=-\left(1-\dfrac{r^{2}-r^{2}f(r)}{\rho^{2}}\right)dt^{2}+\dfrac{\rho^{2}}{\Delta}dr^{2}-\dfrac{2a\sin^{2}\theta\left(r^{2}-r^{2}f(r)\right)}{\rho^{2}}dtd\phi+\rho^{2}d\theta^{2}+\dfrac{\Sigma\sin^{2}\theta}{\rho^{2}}d\phi^{2}$$$$
=-\left(1-\dfrac{2Mr-\frac{Q_{m}^{2k}}{r^{2k-2}}}{\rho^{2}}  \right)dt^{2}+\dfrac{\rho^{2}}{\Delta}dr^{2}-\dfrac{2a\sin^{2}\theta\left(2Mr-\frac{Q_{m}^{2k}}{r^{2k-2}}\right)}{\rho^{2}}dtd\phi+\rho^{2}d\theta^{2}+\dfrac{\Sigma\sin^{2}\theta}{\rho^{2}}d\phi^{2}.
\label{25}
\end{equation}

Where $\rho^{2}=r^{2}+a^{2}\cos^{2}\theta$, $\Sigma=(r^{2}+a^{2})^{2}-a^{2}\Delta(r)\sin^{2}\theta$, $\Delta=r^{2}-2Mr+\frac{Q_{m}^{2k}}{r^{2k-2}}+a^{2}$. If we consider a no-hair black hole, that is, $Q_{m}=0$, then the metric (2.25) degrades to a Kerr black hole, and $a$ can be interpreted as the black hole spin. When $Q_{m}\neq 0$ and $k>1$, it is a short-hair black hole in rotation situation.

When $k=\frac{3}{2}$, $\Delta=r^{2}-2Mr+\frac{Q_{m}^{3}}{r}+a^{2}=0$ will determine the structure of the black hole event horizon. By adjusting the range of values of $M$, $Q_{m}$ and $a$, the number of the black hole event horizon may be $1$, $2$ or $3$. This is caused by the introduction of the short hair. Then we analyse the specific properties of each case. From $\Delta=0$, it can be found that the event horizon of the black hole should satisfy the following equation 

\begin{equation}
r^{3}-2Mr^{2}+a^{2}r+Q_{m}^{3}=0
\label{26}
\end{equation}

The discriminant of the root is

\begin{equation}
\Delta=\left(\dfrac{a^{2}}{3}-\dfrac{4M^{2}}{9}\right)^{3}+\left(\dfrac{Q_{m}^{3}}{2}-\dfrac{8M^{3}}{27}+\dfrac{Ma^{2}}{3}\right)^{2}
\label{27}
\end{equation}

When $\Delta>0$, equation (\ref{26}) has only one real root, so the black hole has only one event horizon, the radius of which is 

\begin{equation}
r=\dfrac{2M}{3}+\sqrt[3]{\dfrac{8M^{3}}{27}-\dfrac{Q_{m}^{3}}{2}-\dfrac{Ma^{2}}{3}+\sqrt{\Delta}}+\sqrt[3]{\dfrac{8M^{3}}{27}-\dfrac{Q_{m}^{3}}{2}-\dfrac{Ma^{2}}{3}-\sqrt{\Delta}}  
\label{28}
\end{equation}

When $\Delta=0$, equation (\ref{26}) has three real roots, at least two of which are equal, so the event horizons of the black hole are  

\begin{equation}
r_{1}=\dfrac{2M}{3}-2\sqrt[3]{\dfrac{Q_{m}^{3}}{2}-\dfrac{8M^{3}}{27}+\dfrac{Ma^{2}}{3}}
\label{29}
\end{equation}

\begin{equation}
r_{2}=r_{3}=\dfrac{2M}{3}+\sqrt[3]{\dfrac{Q_{m}^{3}}{2}-\dfrac{8M^{3}}{27}+\dfrac{Ma^{2}}{3}}
\label{30}
\end{equation}

When $\Delta<0$, equation (\ref{26}) has three unequal real roots, that is, the black hole has three different event horizons, and their values are  

\begin{equation}
r_{1}=\dfrac{2M}{3}+2\sqrt{\dfrac{4M^{2}}{9}-\dfrac{a^{2}}{3}}\cos\left[\dfrac{1}{3}\arccos\left(\dfrac{\left(\dfrac{8M^{3}}{27}-\dfrac{Ma^{2}}{3}-\dfrac{Q_{m}^{3}}{2}\right)\sqrt{\dfrac{4}{9}M^{2}-\dfrac{1}{3}a^{2}}}{\left(\dfrac{1}{3}a^{2}-\dfrac{4}{9}M^{2}\right)^{2}}\right)\right]
\label{31}
\end{equation}

\begin{equation}
r_{2}=\dfrac{2M}{3}+2\sqrt{\dfrac{4M^{2}}{9}-\dfrac{a^{2}}{3}}\cos\left[\dfrac{1}{3}\arccos\left(\dfrac{\left(\dfrac{8M^{3}}{27}-\dfrac{Ma^{2}}{3}-\dfrac{Q_{m}^{3}}{2}\right)\sqrt{\dfrac{4}{9}M^{2}-\dfrac{1}{3}a^{2}}}{\left(\dfrac{1}{3}a^{2}-\dfrac{4}{9}M^{2}\right)^{2}}\right)+\dfrac{2}{3}\pi\right]
\label{32}
\end{equation}

\begin{equation}
r_{3}=\dfrac{2M}{3}+2\sqrt{\dfrac{4M^{2}}{9}-\dfrac{a^{2}}{3}}\cos\left[\dfrac{1}{3}\arccos\left(\dfrac{\left(\dfrac{8M^{3}}{27}-\dfrac{Ma^{2}}{3}-\dfrac{Q_{m}^{3}}{2}\right)\sqrt{\dfrac{4}{9}M^{2}-\dfrac{1}{3}a^{2}}}{\left(\dfrac{1}{3}a^{2}-\dfrac{4}{9}M^{2}\right)^{2}}\right)+\dfrac{4}{3}\pi\right]
\label{33}
\end{equation}

Through calculations, we find that the structure of the event horizon of the short-hair black hole is particularly complicated. $\Delta$ determines the number of black hole event horizons. Here, the mass of the black hole is taken as the unit, i.e., $M=1$. When the black hole spin $0\leqslant a \leqslant 1$, the critical value condition of $Q_{m}$ derived from $\Delta=0$ is 

\begin{equation}
Q_{m}^{3}=2\left(\dfrac{4}{9}-\dfrac{a^{2}}{3}\right)^{\frac{3}{2}}+\dfrac{16}{27}-\dfrac{2}{3}a^{2}
\label{34}
\end{equation}

When $\Delta>0$, i.e., $Q_{m}^{3}<2(\frac{4}{9}-\frac{a^{2}}{3})^{\frac{3}{2}}+\frac{16}{27}-\frac{2a^{2}}{3}$, only $Q_{m}<0$ can meet this situation, which indicates that the short-hair black hole has no event horizon, so it is not the case we consider. When $\Delta=0$, the value range of $Q_{m}$ is $0 \leqslant Q_{m} \leqslant \frac{2}{3}\times 4^{\frac{1}{3}}$. When $\Delta<0$, the value range is $Q_{m}>\frac{2}{3}\times 4^{\frac{1}{3}}$. In general, when $0 \leqslant Q_{m} \leqslant \frac{2}{3}\times 4^{\frac{1}{3}}$, the short-hair black hole has three event horizons, two of which are equal. From a physical point of view, there are only two event horizons, which are modifications of Kerr black hole. When $Q_{m}>\frac{2}{3}\times 4^{\frac{1}{3}}$, the short-hair black hole has three different event horizons, which indicates that the appearance of the short-hair makes the black hole appear a new event horizon, and essentially changes the structure of the black hole event horizon. 

As we all know, AdS-Kerr black hole has three event horizons $(r_{+},r_{-},r_{\Lambda})$. The radius of the event horizon generated by the introduction of cosmological constant $\Lambda$ is very large, namely $r_{\Lambda}>>r_{+}$, while the radius of the event horizon generated by the short hair is close to $r_{+}$. Since the physical effects of $r_{+}$ are relatively easy to measure, the corresponding effects caused by the short hair are also relatively easy to detect, which makes the third event horizon of the short-hair black hole interesting.

\section{Geodesic equations of photons and analytic solution}
\label{3}

Next, we derive the geodesic equations of photons based on the rotational space-time metrics (\ref{7}) and (2.25), and obtain the analytical solution. We will use the Hamilton-Jacobi equation to calculate. In this method, Carter et al. introduced a new integral constant and obtained analytical solutions of the geodesic equations by using variable separation approach\citep{1968PhRv..174.1559C,Chandrasekhar:1985kt}. Now, let's introduce the main process. For a rotational black hole, the Hamilton-Jacobi equation for the test particle is of the following form 

\begin{equation}
\dfrac{\partial S}{\partial \sigma}=-\dfrac{1}{2}g^{\mu\nu}\dfrac{\partial S}{\partial x^{\mu}}\dfrac{\partial S}{\partial x^{\nu}},
\label{35}
\end{equation}

where $S$ is the Jacobi action and $\sigma$ is the affine parameter of a geodesic. In order to be able to separate variables from geodesic equations, the action $S$ should have the form  

\begin{equation}
S=\dfrac{1}{2}m^{2}\sigma-Et+\mathcal{L}\phi+S_{r}(r)+S_{\theta}(\theta).
\label{36}
\end{equation}

$m$ is the mass of the test particle, and for photons, $m=0$. $E$ and $\mathcal{L}$ are the two initial integrals of the motion equation, corresponding to energy and angular momentum respectively. $S_{r}(r)$ and $S_{\theta}(\theta)$ are radial and angular functions respectively. Due to the introduction of these conserved quantities and unknown functions, the geodesic equations can be simplified to four component equations. These results can be achieved in the following ways. By substituting the action (\ref{36}) into the Hamilton-Jacobi equation, equation (\ref{35}) can be reduced to such equations  

\begin{equation}
\begin{aligned}
& \rho^{2}\dfrac{dt}{d\sigma}=\dfrac{r^{2}+a^{2}}{\Delta(r)}\left(E(r^{2}+a^{2})-a\mathcal{L} \right)-a\left(aE\sin^{2}\theta-\mathcal{L} \right),  \\
& \rho^{2}\dfrac{dr}{d\sigma}=\sqrt{R}, ~~~~~~~~ \rho^{2}\dfrac{d\theta}{d\sigma}=\sqrt{H},  \\
& \rho^{2}\dfrac{d\phi}{d\sigma}=\dfrac{a}{\Delta}\left(E(r^{2}+a^{2})-a\mathcal{L} \right)-\left(aE-\dfrac{\mathcal{L}}{\sin^{2}\theta} \right),
\end{aligned}
\label{37}
\end{equation}

Since unknown functions $S_{r}(r)$ and $S_{\theta}(\theta)$ are set in the action $S$, we need to introduce functions $R$ and $H$, whose relationship with other initial integrals is  

\begin{equation}
R=\left(E(r^{2}+a^{2})-a\mathcal{L} \right)^{2}-\Delta\left(m^{2}r^{2}+(aE-\mathcal{L})^{2}+K \right)
\label{38}
\end{equation}

\begin{equation}
H=K-\left(\dfrac{\mathcal{L}^{2}}{\sin^{2}\theta}-a^{2}E^{2} \right)\cos^{2}\theta
\label{39}
\end{equation}

Where $K$ is the constant introduced by Carter et al., which is later called Carter constant. By introducing Carter constant, we can separate variables from the motion equation, and (\ref{37}) $\sim$ (\ref{39}) are complete geodesic equations. In this work, we calculate the motion of the test particle on the equatorial plane of the black hole, so $\theta=\frac{\pi}{2}$. In addition, in the process of calculating the black hole shadow, the test particle is generally considered as photon, whose static mass is $0$. The motion of photons near black holes is very complex and is discussed in detail in the relevant classical textbooks. The most obvious feature is that at a certain radius close to the event horizon of the black hole, the motion of the photon in this critical radius is very different from that outside. There is no stable motion orbit of the photon within this critical radius (the region satisfies the condition $R^{''}<0, R^{'}=dR/dr$ ), but there exist stable motion orbit outside this radius(the region satisfies the condition $R^{''}>0$). The critical orbit should satisfy $R=0$ and $\frac{dR}{dr}=0$. In order to better introduce the coordinates of the shape of the black hole shadow, we first introduce the parameter pair $(\xi,\eta)$, so that $\mathcal{L}, K, E$ can be expressed. By substituting $R$ and considering $\xi=\frac{\mathcal{L}}{E}$ and $\eta=\frac{K}{E^{2}}$, $\xi$ and $\eta$ can be reduced to  

$  $

For Hairy black hole

\begin{equation}
\xi=
{\dfrac{1}{a}\left[\left(2\alpha_{0}-\dfrac{\alpha_{0}r}{M-\frac{\alpha_{0}L}{2}} \right)\exp\left(\dfrac{-r}{M-\frac{\alpha_{0}L}{2}} \right)+2-\dfrac{2M}{r} \right]^{-1}}
\left\lbrace  \left(r^{2}+a^{2}\right)\left[\left(2\alpha_{0}-\dfrac{\alpha_{0}r}{M-\frac{\alpha_{0}L}{2}} \right) \right.\right. $$$$
\left.\left. \times\exp\left(\dfrac{-r}{M-\frac{\alpha_{0}L}{2}} \right) 
 +2-\dfrac{2M}{r} \right] -4\left[a^{2}+r^{2}\left(1+\alpha_{0}\exp\left(\dfrac{-r}{M-\frac{\alpha_{0}L}{2}} \right) \right)-2Mr \right] \right \rbrace   
\label{40}
\end{equation}

\begin{equation}
\eta=
{\dfrac{1}{a^{2}}\left[\left(2\alpha_{0}-\dfrac{\alpha_{0}r}{M-\frac{\alpha_{0}L}{2}} \right)\exp\left(\dfrac{-r}{M-\frac{\alpha_{0}L}{2}} \right)+2-\dfrac{2M}{r} \right]^{-1}}
\left\lbrace  16Ma^{2}-\dfrac{8a^{2}\alpha_{0}r^{2}}{M-\frac{\alpha_{0}L}{2}}   \right.  $$$$
\left. \times\exp\left(\dfrac{-r}{M-\frac{\alpha_{0}L}{2}} \right)  -r^{3}\left[2-\dfrac{6M}{r}+\left(2\alpha_{0}+\dfrac{\alpha_{0}r}{M-\frac{\alpha_{0}L}{2}} \right)\exp\left(\dfrac{-r}{M-\frac{\alpha_{0}L}{2}} \right) \right]^{2}    \right\rbrace
\label{41}
\end{equation}

$ $

For short-hair black hole

\begin{equation}
\xi=
\dfrac{\left(r^{2}+a^{2}\right) \left[2-\dfrac{2M}{r}+\dfrac{2Q_{m}^{2k}(1-k)}{r^{2k}}\right] -4\left[r^{2}-2Mr+\dfrac{Q_{m}^{2k}}{r^{2k-2}}+a^{2}\right]} {a\left[ 2-\dfrac{2M}{r}+\dfrac{2Q_{m}^{2k}(1-k)}{r^{2k}} \right]}
\label{42}
\end{equation}

\begin{equation}
\eta=
\dfrac{16a^{2}\left(M-\dfrac{kQ_{m}^{2k}}{r^{2k-1}}\right) -4r^{3}\left[1-\dfrac{3M}{r}+\dfrac{Q_{m}^{2k}(k+1)}{r^{2k}}\right]^{2}} {a^{2}\left[ 2-\dfrac{2M}{r}+\dfrac{2Q_{m}^{2k}(1-k)}{r^{2k}} \right]}.
\label{43}
\end{equation}

\section{The no-hair theorem and the shape of black hole shadows}
\label{4}

\subsection{The shape of black hole shadows}

Once the geodesic of the photon is known, we can calculate the motion of the photon as measured by an observer anywhere. In general, we assume that the earth is infinitely far away from the black hole and the observer is a mass point, which makes the calculation very convenient. In the calculation of black hole shadows, physicists introduce the celestial coordinate system, which is a two-dimensional coordinate system\citep{2009PhRvD..80b4042H}, and its relationship with the BL coordinates of the black hole is  

\begin{equation}
\alpha=\lim\limits_{r_{0}\to\infty} \left(-r_{0}^{2}\sin\theta_{0}\dfrac{d\phi}{dr} \right) 
\label{44}
\end{equation}

\begin{equation}
\beta=\lim\limits_{r_{0}\to\infty} \left(r_{0}^{2}\dfrac{d\theta}{dr} \right) 
\label{45}
\end{equation}

where $\theta_{0}$ is the angle between the line from the earth to the black hole and the axis of rotation of the black hole, and $r_{0}$ is the distance between the earth and the black hole. In a celestial coordinate system, photons emitted from the vicinity of the black hole correspond to the coordinates $(\alpha, \beta)$ one by one, so that the geometry of the black hole shadow to be understood by pushing back the edge of the geodesic motion of the photon by $(\alpha, \beta)$. Through calculation, we get the expressions of $\alpha$ and $\beta$ as  

$ $

For Hairy black hole

\begin{equation}
\alpha= -\dfrac{\xi}{\sin\theta}\mid_{\theta=\frac{\pi}{2}}=-\xi
=  $$$$
-{\dfrac{1}{a}\left[\left(2\alpha_{0}-\dfrac{\alpha_{0}r}{M-\frac{\alpha_{0}L}{2}} \right)\exp\left(\dfrac{-r}{M-\frac{\alpha_{0}L}{2}} \right)+2-\dfrac{2M}{r} \right]^{-1}}
\left\lbrace  \left(r^{2}+a^{2}\right)\left[\left(2\alpha_{0}-\dfrac{\alpha_{0}r}{M-\frac{\alpha_{0}L}{2}} \right) \right.\right. $$$$
\left.\left. \times\exp\left(\dfrac{-r}{M-\frac{\alpha_{0}L}{2}} \right) 
 +2-\dfrac{2M}{r} \right] -4\left[a^{2}+r^{2}\left(1+\alpha_{0}\exp\left(\dfrac{-r}{M-\frac{\alpha_{0}L}{2}} \right) \right)-2Mr \right] \right \rbrace   
\label{46}
\end{equation}

\begin{equation}
\beta=\pm\sqrt{\eta+a^{2}\cos^{2}\theta-\xi^{2}\cot^{2}\theta}\mid_{\theta=\frac{\pi}{2}}=\pm\sqrt{\eta}=   $$$$
\pm{\dfrac{1}{a^{2}}\left[\left(2\alpha_{0}-\dfrac{\alpha_{0}r}{M-\frac{\alpha_{0}L}{2}} \right)\exp\left(\dfrac{-r}{M-\frac{\alpha_{0}L}{2}} \right)+2-\dfrac{2M}{r} \right]^{-1}}
\left\lbrace  16Ma^{2}-\dfrac{8a^{2}\alpha_{0}r^{2}}{M-\frac{\alpha_{0}L}{2}}   \right.  $$$$
\left. \times\exp\left(\dfrac{-r}{M-\frac{\alpha_{0}L}{2}} \right)  -r^{3}\left[2-\dfrac{6M}{r}+\left(2\alpha_{0}+\dfrac{\alpha_{0}r}{M-\frac{\alpha_{0}L}{2}} \right)\exp\left(\dfrac{-r}{M-\frac{\alpha_{0}L}{2}} \right) \right]^{2}    \right\rbrace
\label{47}
\end{equation}

$ $

For short-hair black hole

\begin{equation}
\alpha=-\xi=-
\dfrac{\left(r^{2}+a^{2}\right) \left[2-\dfrac{2M}{r}+\dfrac{2Q_{m}^{2k}(1-k)}{r^{2k}}\right] -4\left[r^{2}-2Mr+\dfrac{Q_{m}^{2k}}{r^{2k-2}}+a^{2}\right]} {a\left[ 2-\dfrac{2M}{r}+\dfrac{2Q_{m}^{2k}(1-k)}{r^{2k}} \right]}
\label{48}
\end{equation}

\begin{equation}
\beta=\pm\sqrt{\eta}=
\pm\dfrac{16a^{2}\left(M-\dfrac{kQ_{m}^{2k}}{r^{2k-1}}\right) -4r^{3}\left[1-\dfrac{3M}{r}+\dfrac{Q_{m}^{2k}(k+1)}{r^{2k}}\right]^{2}} {a^{2}\left[ 2-\dfrac{2M}{r}+\dfrac{2Q_{m}^{2k}(1-k)}{r^{2k}} \right]}
\label{49}
\end{equation}

\begin{figure}[htbp]
  \centering
   \includegraphics[scale=0.3]{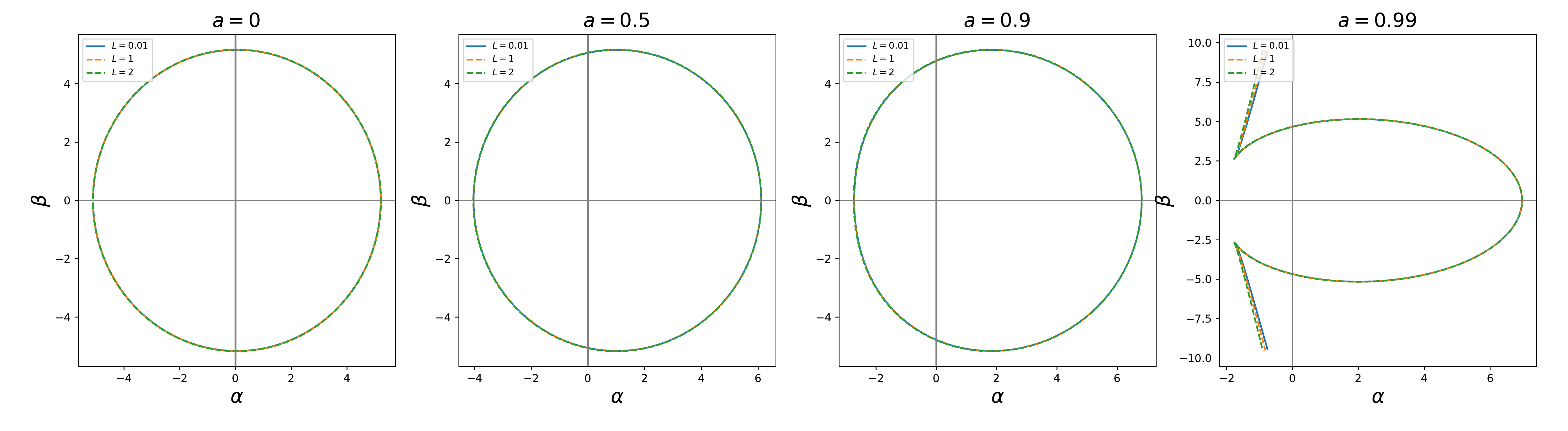}
   \includegraphics[scale=0.3]{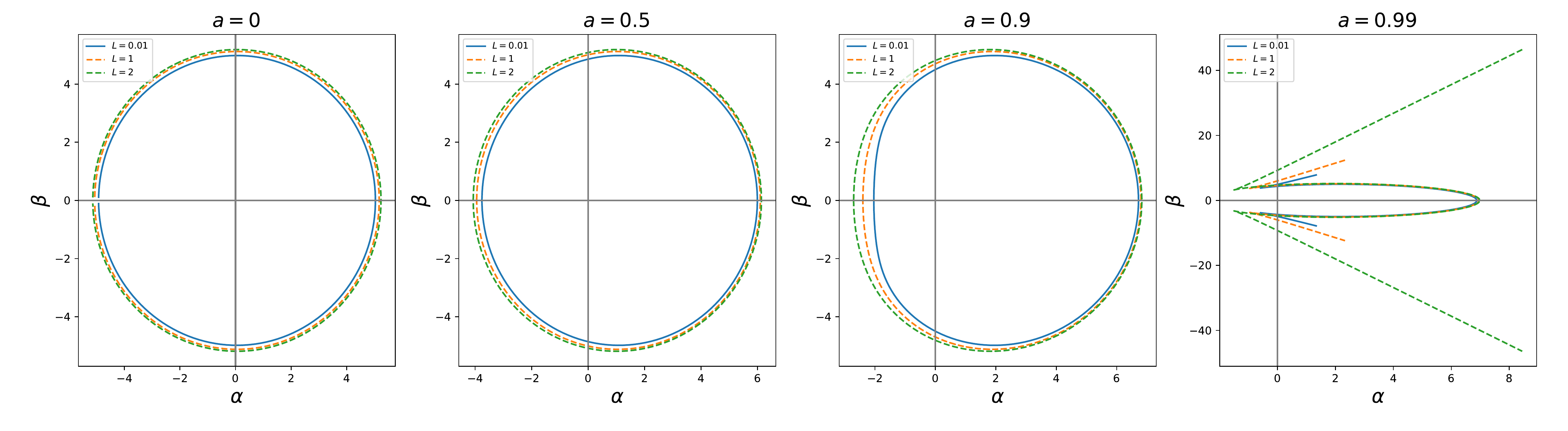}
   \includegraphics[scale=0.3]{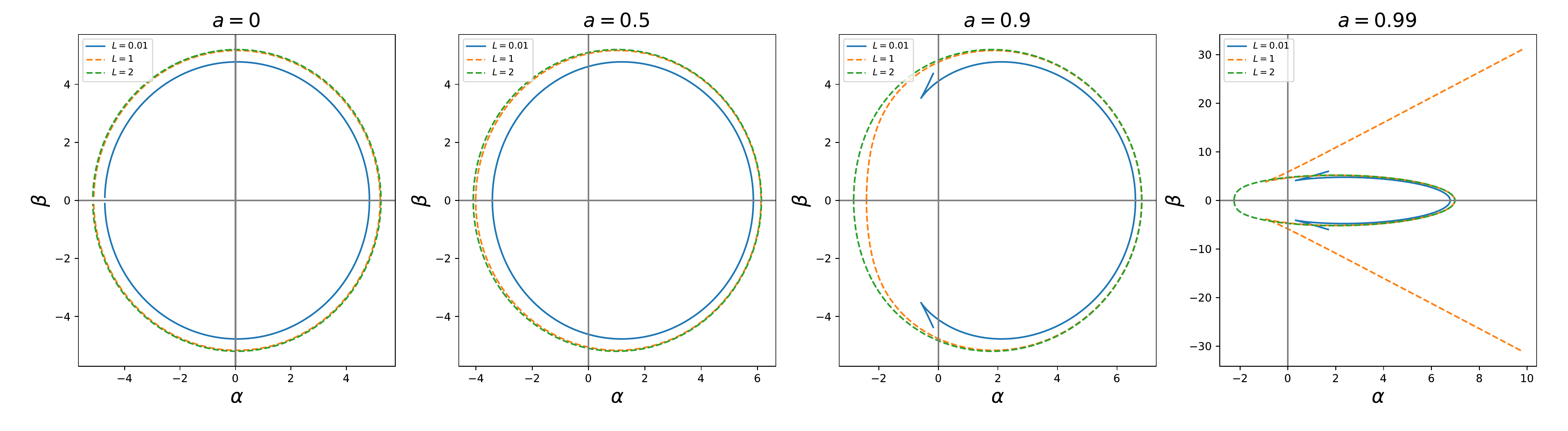}
   \caption{The shadow shape of hairy black hole under different model parameters (Model 1). From left to right represents the process of increasing the spin of the black hole, which is $a=0, 0.5, 0.9, 0.99$ respectively. From top to bottom represents the process of increasing deformation parameter, which is $\alpha_{0}=0.1, 0.5, 0.9$ respectively. Curves of various colors correspond to different $L$ values.  }
  \label{shadow_type1_Kerr_alpha0}
\end{figure}

\begin{figure}[htbp]
  \centering
   \includegraphics[scale=0.29]{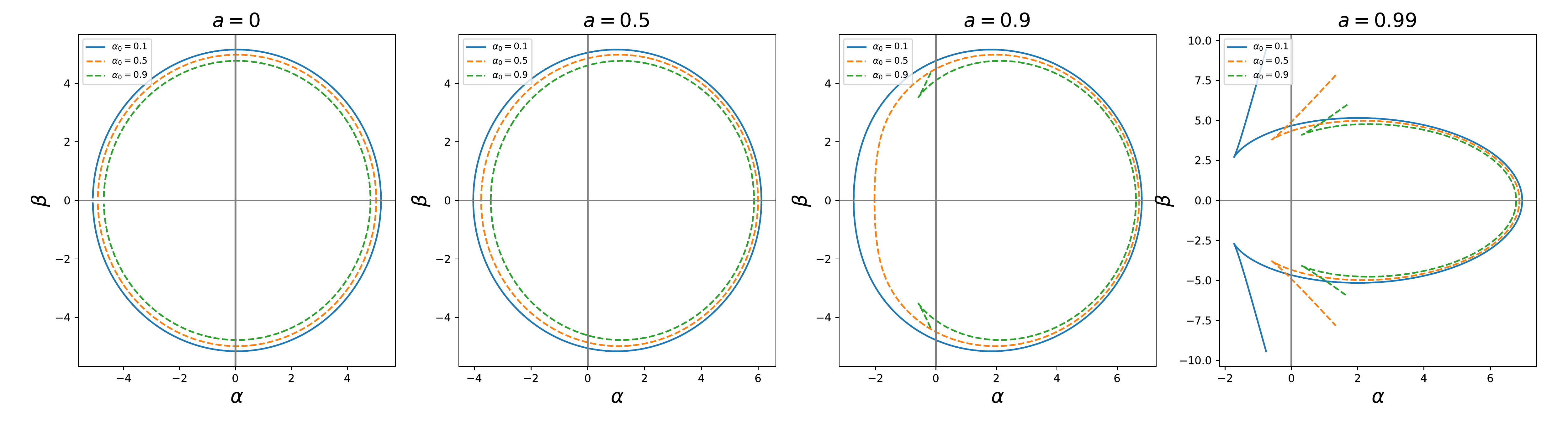}
   \includegraphics[scale=0.29]{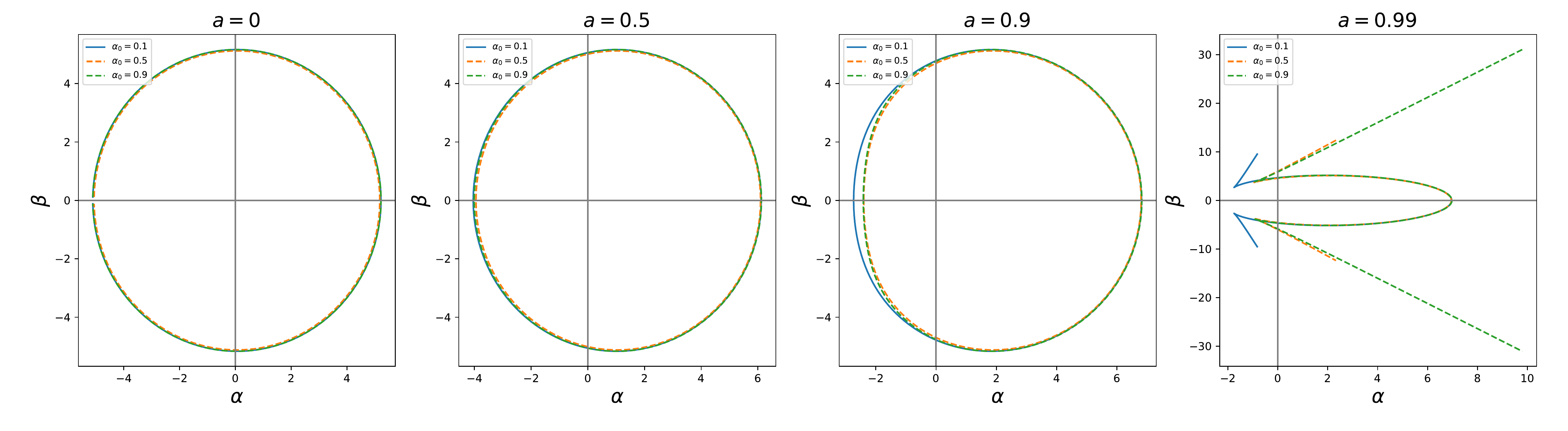}
   \includegraphics[scale=0.29]{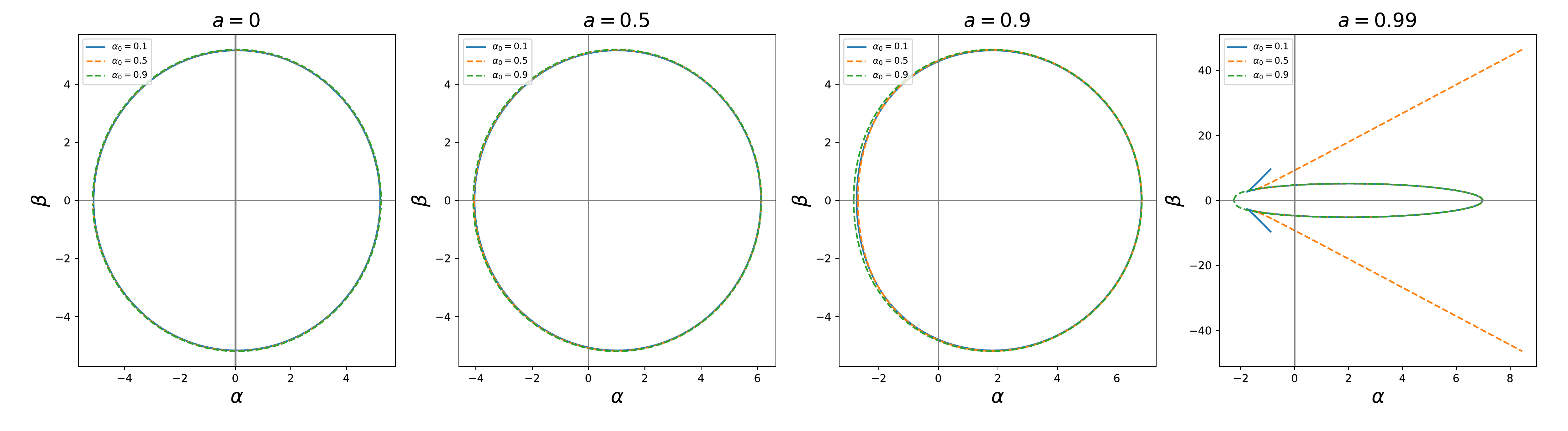}
   \caption{The shadow shape of hairy black hole under different model parameters (Model 1). From left to right represents the process of increasing the spin of the black hole, which is $a=0, 0.5, 0.9, 0.99$ respectively. From top to bottom represents the process of increasing parameter $L$, which is $L=0.01, 1, 2$ respectively. Curves of various colors correspond to different $\alpha_{0}$ values.}
  \label{shadow_type1_Kerr_L}
\end{figure}

\begin{figure}[htbp]
  \centering
   \includegraphics[scale=0.3]{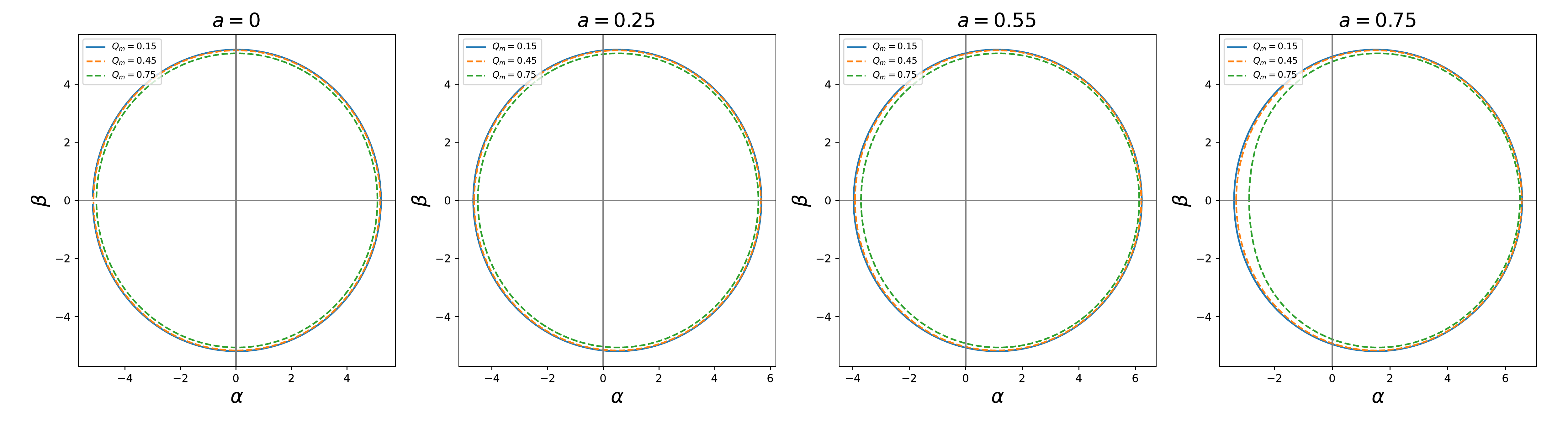}
   \includegraphics[scale=0.3]{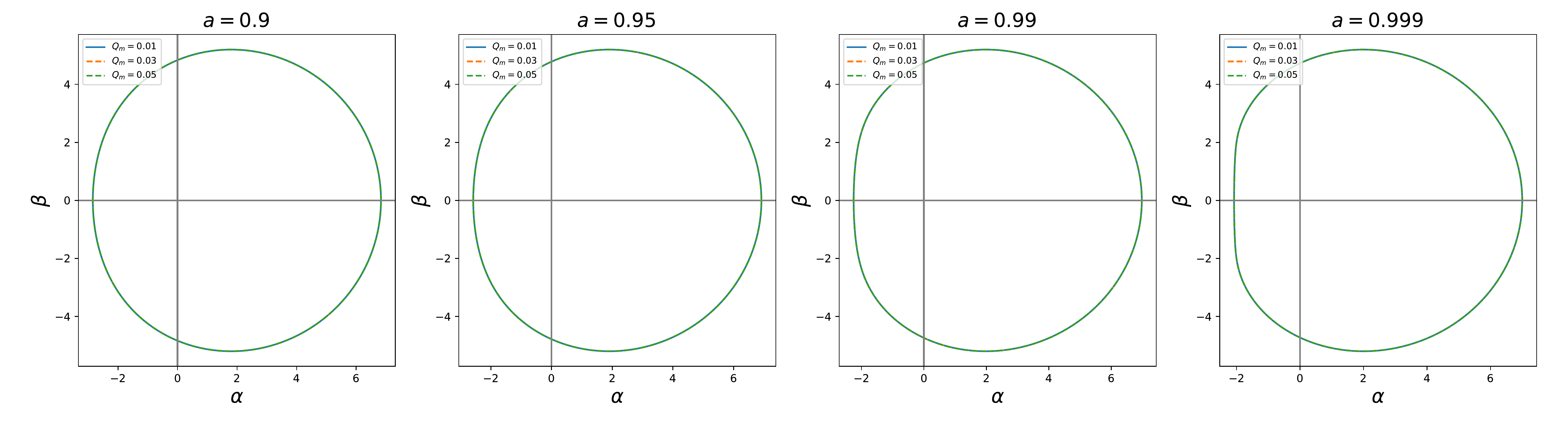}
   \includegraphics[scale=0.3]{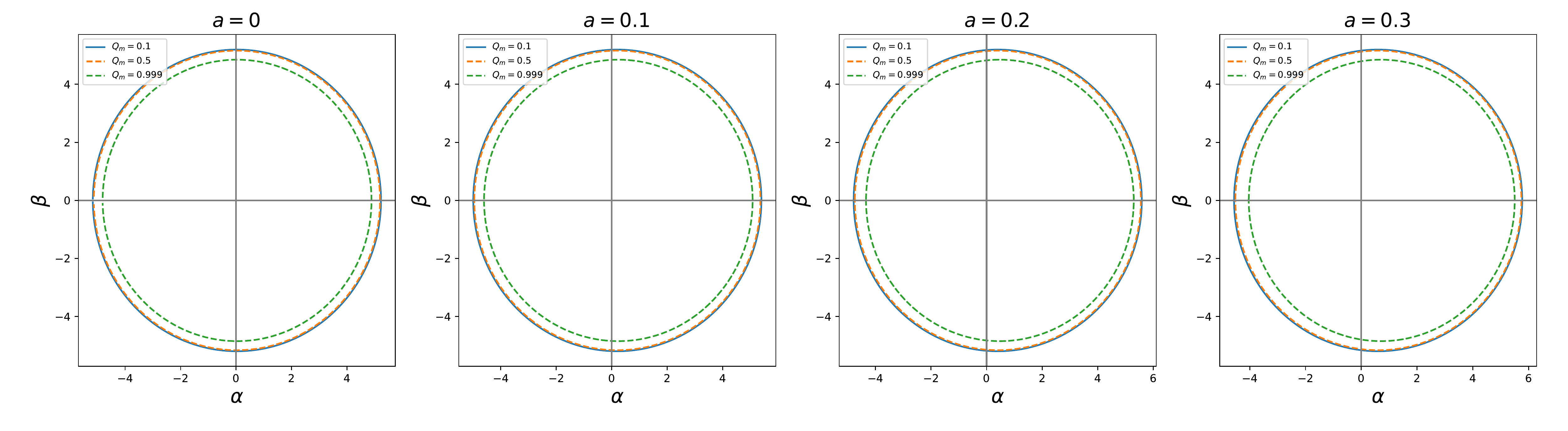}
   \caption{The shadow shape of short-hair black hole under different model parameters (Model 2). From left to right is the process of increasing the black hole spin $a$. Curves of various colors correspond to different values of intensity parameter $Q_{m}$. Here $k=1.5$.}
  \label{shadow_type2_k}
\end{figure}

Using expressions (\ref{46}) $\thicksim$ (\ref{49}), the shape of the black hole shadow can be calculated, so as to make clear the influence of the hairs on the shape of black hole shadow, as shown in figures \ref{shadow_type1_Kerr_alpha0}, \ref{shadow_type1_Kerr_L}, and \ref{shadow_type2_k}. The main results are as follows: 

We obtain the shadow shapes of two kinds of black holes with hairs. In model $1$ (corresponding to figures \ref{shadow_type1_Kerr_alpha0} and \ref{shadow_type1_Kerr_L}), the values of parameters $\alpha_{0}$ and $L$ have a significant influence on the shadow shapes (mainly referring to the size of the shadow boundary here). When the parameter $\alpha_{0}$, which represents the intensity of the scalar hair, increases gradually from $0$, the boundary of the black hole shadow will shrink continuously. When $\alpha_{0}$ increases to a certain critical value, the black hole shadow changes in reverse. On the other hand, the boundary of the black hole shadow will always increase with the increase of the characteristic length $L$. Therefore, the change of $\alpha_{0}$ and $L$ to the black hole shadow is complicated. In Model $2$ (corresponding to figure \ref{shadow_type2_k}), when $k=1.5$, namely the short hair case, if the black hole spin $a=0$, the shape of the shadow boundary of the black hole is a standard circle. It can be found that when the charge $Q_{m}$ of the short hair keeps increasing, the shadow boundary of the black hole will decrease monotonically, but its shape is all circle. As the black hole spin $a$ increases, the shape of the black hole shadow is distorted. We find that the change of the hairy black hole in model $1$ to the shadow is basically consistent with the case $\varepsilon>0$ in reference \citep{2021JCAP...09..028K}, but the specific changes are different. The change of the short-hair black hole in model $2$ to the shadow corresponds to the case $\varepsilon<0$ in reference \citep{2021JCAP...09..028K}.   

If the effects of various hairs carried by black holes on the metric of black holes are known, it is possible to test them by means of EHT observations. However, the parameter values of the real hairy black hole are smaller than those discussed by us (since the shape of the black hole is mainly contributed by the mass, and the quantum effect should not contribute much to it), which requires us to further improve the resolution of EHT before we can test the no-hair theorem of the black hole.

\subsection{The scale and distortion characteristics of shadows}

From the previous analysis, we can see that the boundary shape of the black hole shadow is a circle only when $a=0$. When the black hole spin $a\neq 0$, the boundary of the black hole shadow will be distorted in different degrees, which makes it necessary to introduce new parameters to accurately describe the shape of the black hole shadow. Generally speaking, scientists introduce the following two parameters to describe the shadow shape. The first parameter is the shadow radius $R_{s}$, which describes the scale of the shadow shape of the black hole. The second parameter is the degree of distortion of the shadow ($\delta_{s}$), which describes how far the boundary of the black hole shadow deviates from the circle. According to previous discussions, its definition is as follows:

\begin{equation}
R_{s}=\dfrac{(\alpha_{t}-\alpha_{r})^{2}+\beta_{t}^{2}}{2|\alpha_{r}-\alpha_{t}|},
\label{50}
\end{equation}

\begin{equation}
\delta_{s}=\dfrac{d_{s}}{R_{s}}=\dfrac{|\alpha_{p}-\tilde{\alpha}_{p}|}{R_{s}}.
\label{51}
\end{equation}

\begin{figure}[htbp]
  \centering
   \includegraphics[scale=0.29]{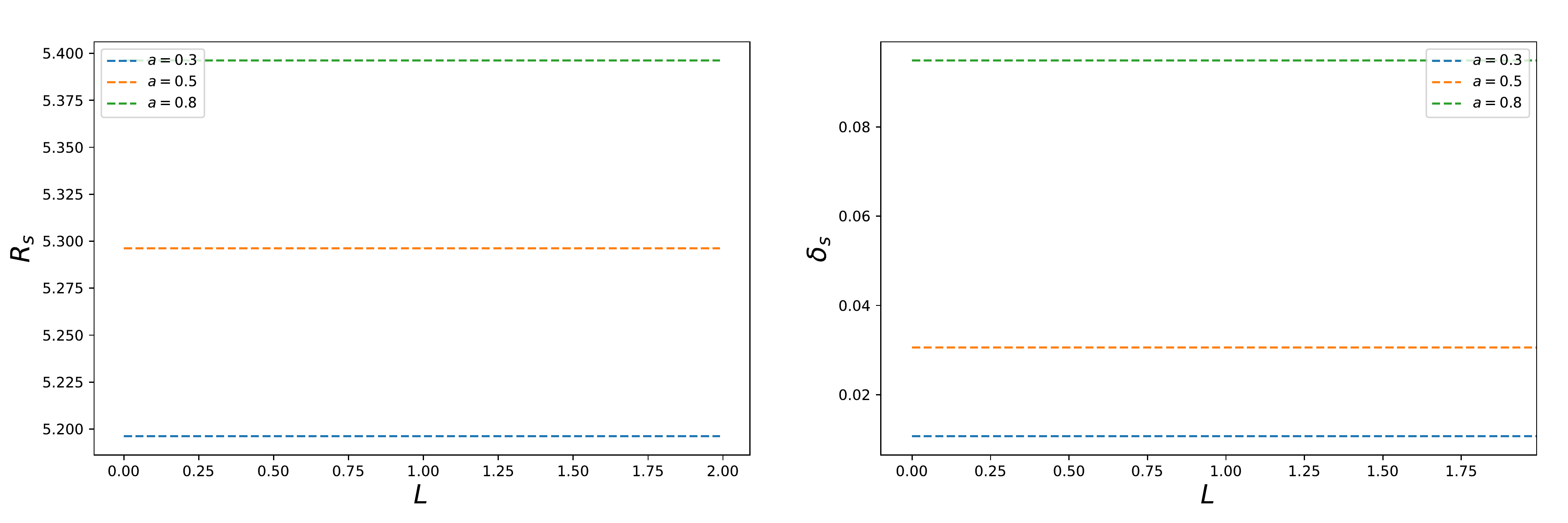}
   \includegraphics[scale=0.29]{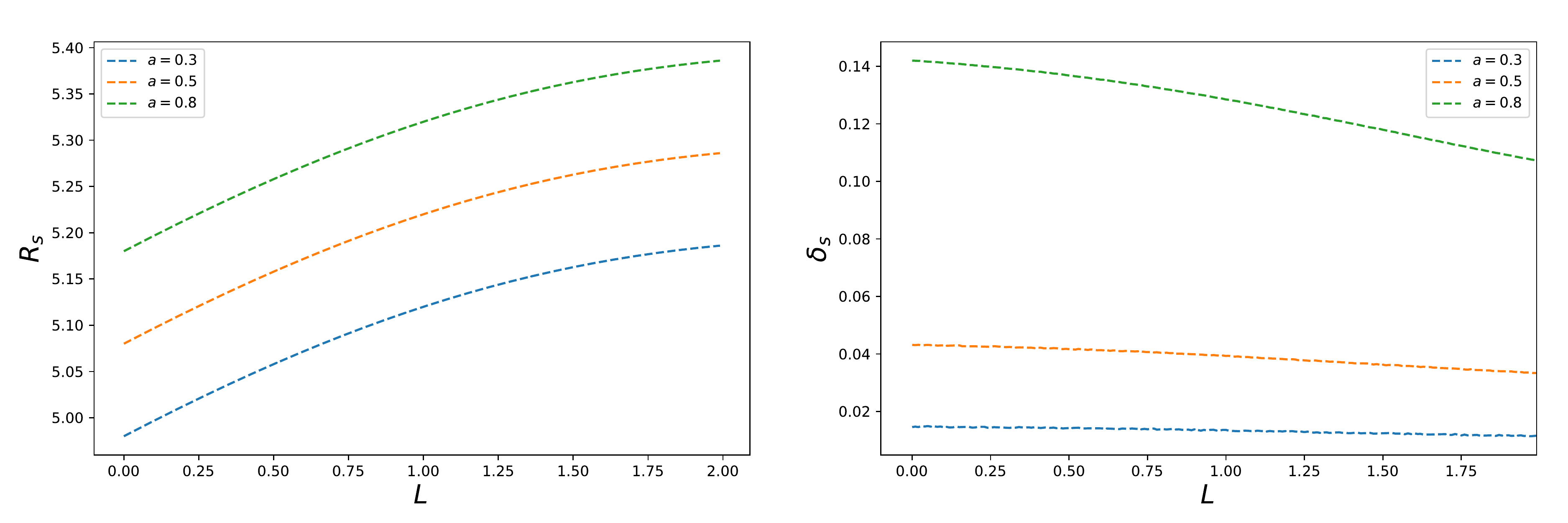}
   \includegraphics[scale=0.29]{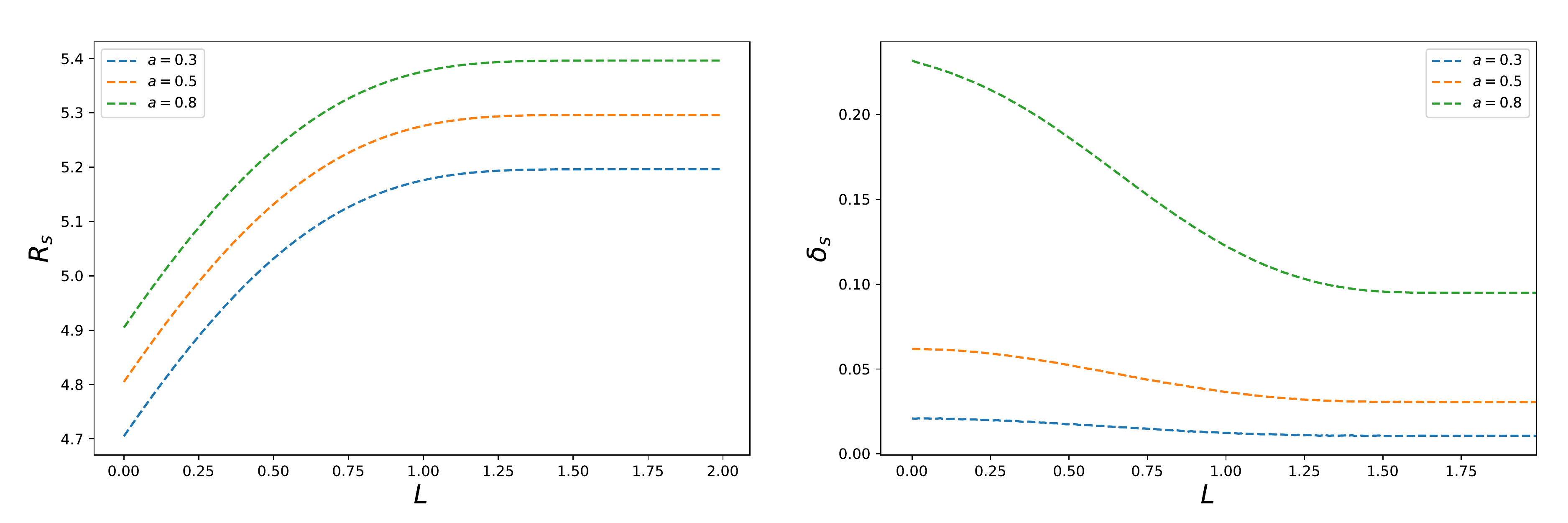}
   \caption{The radius (left) and distortion (right) of the black hole shadow vary with parameter $L$ (Model 1). From top to bottom, they correspond to different deformation parameters $\alpha_{0}=0, 0.5, 1$ respectively. The curves of various colors correspond to different black hole spins $a$.}
  \label{Rs_deltaS_type1_Kerr_alpha0}
\end{figure}

\begin{figure}[htbp]
  \centering
   \includegraphics[scale=0.29]{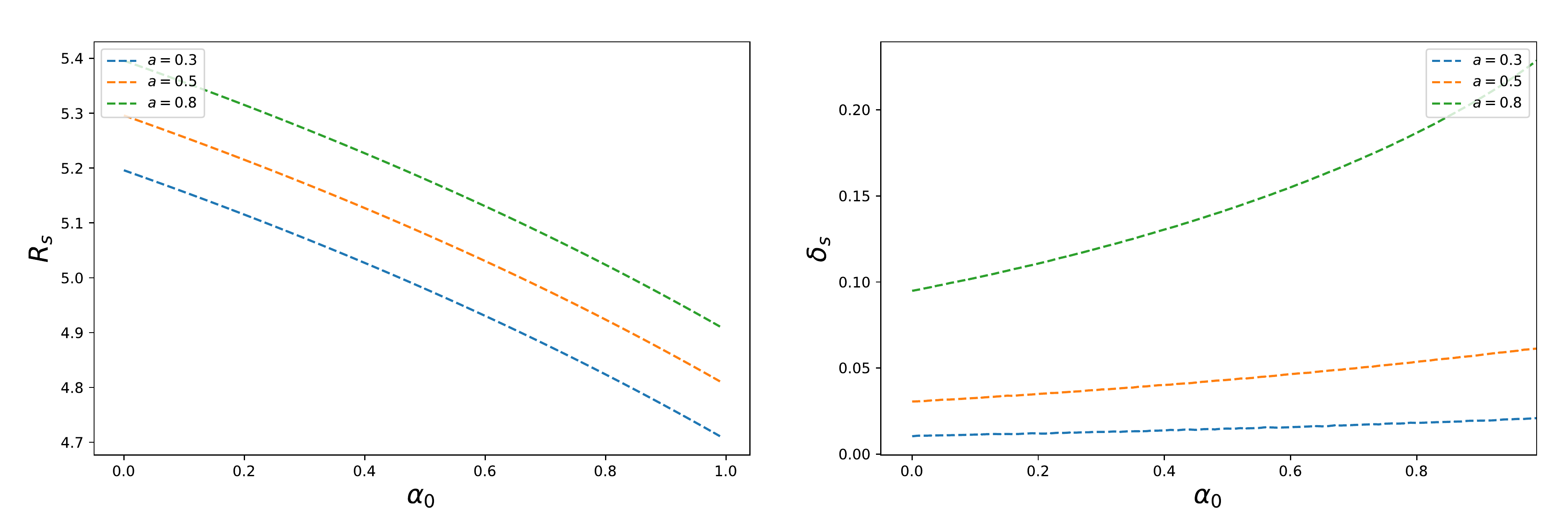}
   \includegraphics[scale=0.29]{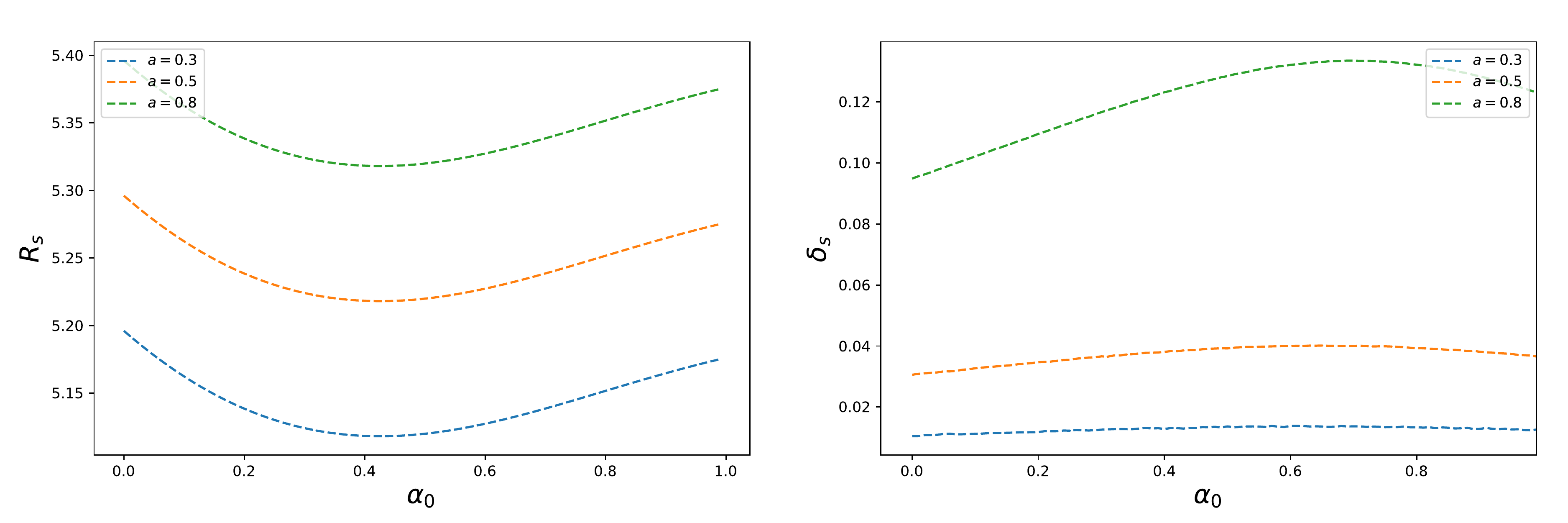}
   \includegraphics[scale=0.29]{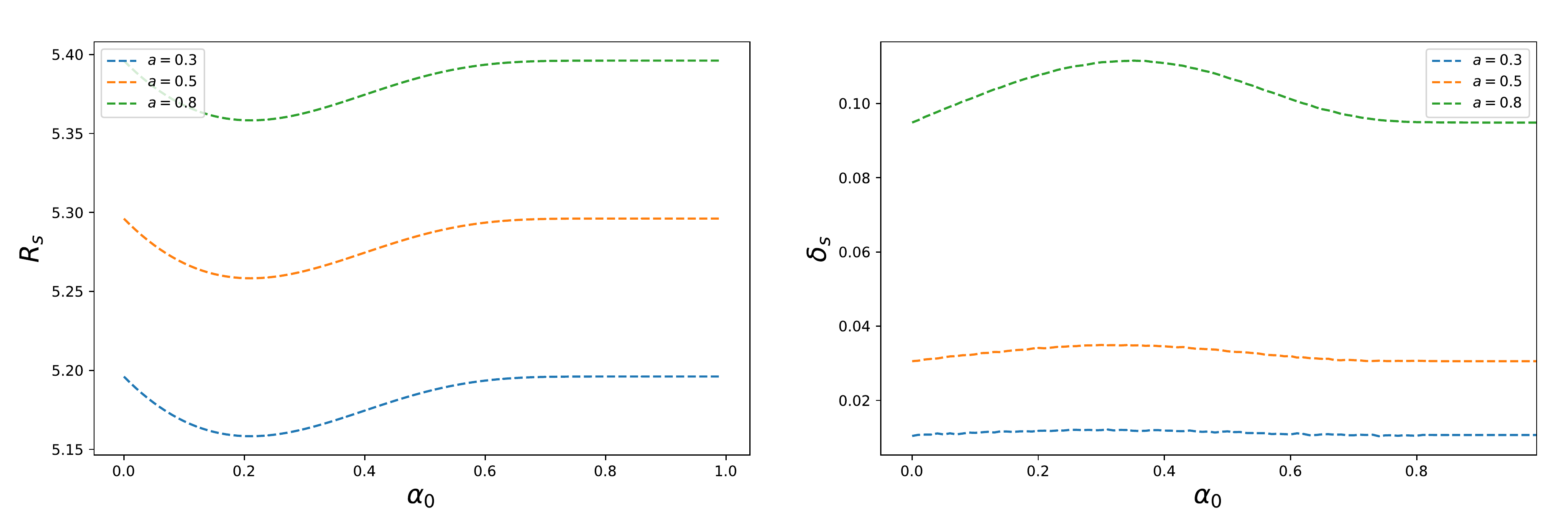}
   \caption{The radius (left) and distortion (right) of the black hole shadow vary with deformation parameter $\alpha_{0}=0$ (Model 1). From top to bottom, they correspond to different parameters $L=0, 1, 2$ respectively. The curves of various colors correspond to different black hole spins $a$.}
  \label{Rs_deltaS_type1_Kerr_L}
\end{figure}

\begin{figure}[htbp]
  \centering
   \includegraphics[scale=0.29]{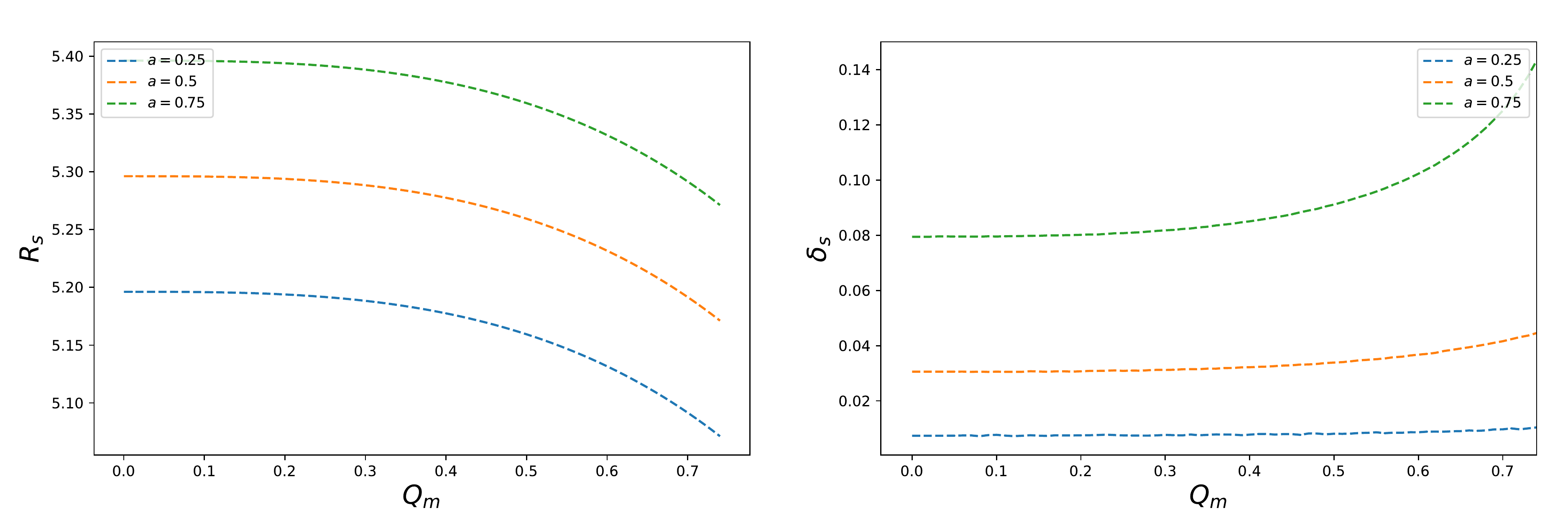}
   \caption{The radius (left) and distortion (right) of the black hole shadow vary with the intensity parameter $Q_{m}$ (Model 2). The curves of various colors correspond to different black hole spins $a$. Here $k=1.5$.}
  \label{Rs_deltaS_type2_k_1}
\end{figure}

The physical meanings of $\alpha_{p}$ and $\tilde{\alpha}_{p}$ are shown in reference \citep{2009PhRvD..80b4042H}. When the black hole spin $a=0$ (i.e. in the case of a spherically symmetric black hole), $\alpha_{p}=\tilde{\alpha}_{p}$, at this time, the black hole shadow is not distorted. When extreme black holes are considered (i.e. $a\rightarrow 1$), $\alpha_{p}-\tilde{\alpha}_{p}$ is at its maximum at this point.  
Through numerical calculation, we find the following results: (1) for model $1$, when Kerr black hole is considered, namely $\alpha_{0}=0$ (no-hair black hole), the change of the scale parameter $L$ hardly affects the value of $R_{s}$. When $\alpha_{0}=0.5$ (i.e. black hole carry hairs), $R_{s}$ increases slowly with the increase of parameter $L$. When $\alpha_{0}=0.8$, that is, more hairs are considered, $R_{s}$ rises rapidly and then changes gently with $L$ (see the left part of Figure \ref{Rs_deltaS_type1_Kerr_alpha0}). When $L=0$, $R_{s}$ decreases with the increase of $\alpha_{0}$. When $L=1$, $R_{s}$ decreases first and then increases with the increase of $\alpha_{0}$. As $L$ increases further, the size of $R_{s}$ varies with $\alpha_{0}$ to show some interesting phenomenon (see the left part of Figure \ref{Rs_deltaS_type1_Kerr_L}).
(2) For model $1$, when Kerr black hole (i.e. no-hair black hole) is considered, the distortion parameter $\delta_{s}$ is a constant function of the scale parameter $L$. When hair black holes are considered, i.e. $\alpha_{0}\neq 0$, $\delta_{s}(L)$ is a decreasing function and decays rapidly at the beginning, while $\delta_{s}$ changes slowly when $L$ is very large (see the right part of Figure \ref{Rs_deltaS_type1_Kerr_alpha0}). In the case that $L$ is constant, $\delta_{s}(\alpha_{0})$ is an increasing function when $L=0$, and the first half is an increasing function and the second half is a decreasing function when $L=1$. Its variation becomes more singular at larger spins of the black hole and $L$ (see the right part of Figure \ref{Rs_deltaS_type1_Kerr_L}).    
(3) For model $2$, i.e., the short-hair black hole, $R_{s}(Q_{m})$ is a monotone decreasing function, $R_{s}(a)$, $\delta_{s}(Q_{m})$ and $\delta_{s}(a)$ are all increasing functions (see Figure \ref{Rs_deltaS_type2_k_1}).  
Since $R_{s}$ and $\delta_{s}$ vary significantly with scalar, quantum or short hairs, it is more likely to be tested in EHT observations.

\section{The rate of energy emission}
\label{5}

When particles or photons pass near a black hole, they are likely to be absorbed by the black hole, so the effect is similar to that of a black body. If the black hole is regarded as a black body, the size of the black hole shadow will be proportional to the high-energy absorption cross section of the particle. In addition, the theoretical calculation of the high-energy absorption cross section ($\sigma_{eim}$) of black holes shows that $\sigma_{eim}$ is very close to a constant, which brings great convenience to our calculation. For a spherically symmetric black hole with hairs (scalar, quantum or short hair), the geometric absorption cross section $\sigma_{eim}$ is approximately equal to the photon sphere of the black hole space-time, i.e., $\sigma_{eim}\approx \pi R_{s}^{2}$\citep{1973PhRvD...7.2807M,2013JCAP...11..063W}. The shape of the black hole shadow is approximately a circle except for the near extreme black hole, so the formula $\sigma_{eim}\approx \pi R_{s}^{2}$ is approximately applicable. Knowing the results of $\sigma_{eim}$, we can well define the rate of energy emission of a rotational black hole as

\begin{equation}
\dfrac{d^{2}E(\omega)}{d\omega dt}=\dfrac{2\pi^{2}\omega^{3}\sigma_{eim}}{e^{\frac{\omega}{T}}-1}.
\label{52}
\end{equation}

Here, $\omega$ is the frequency of the particle and $T$ is the Hawking temperature corresponding to the event horizon of the black hole with hairs, whose mathematical form is directly determined by the metric coefficients of the black hole ($T=\lim\limits_{\theta\to 0,r\to r_{+}}\frac{1}{2\pi\sqrt{g_{rr}}}\frac{\partial\sqrt{g_{tt}}}{\partial r}$). Using Python for numerical calculation, we get the relationship between the rate of energy emission and frequency $\omega$, and then get the following results.

\begin{figure}[htbp]
  \centering
   \includegraphics[scale=0.29]{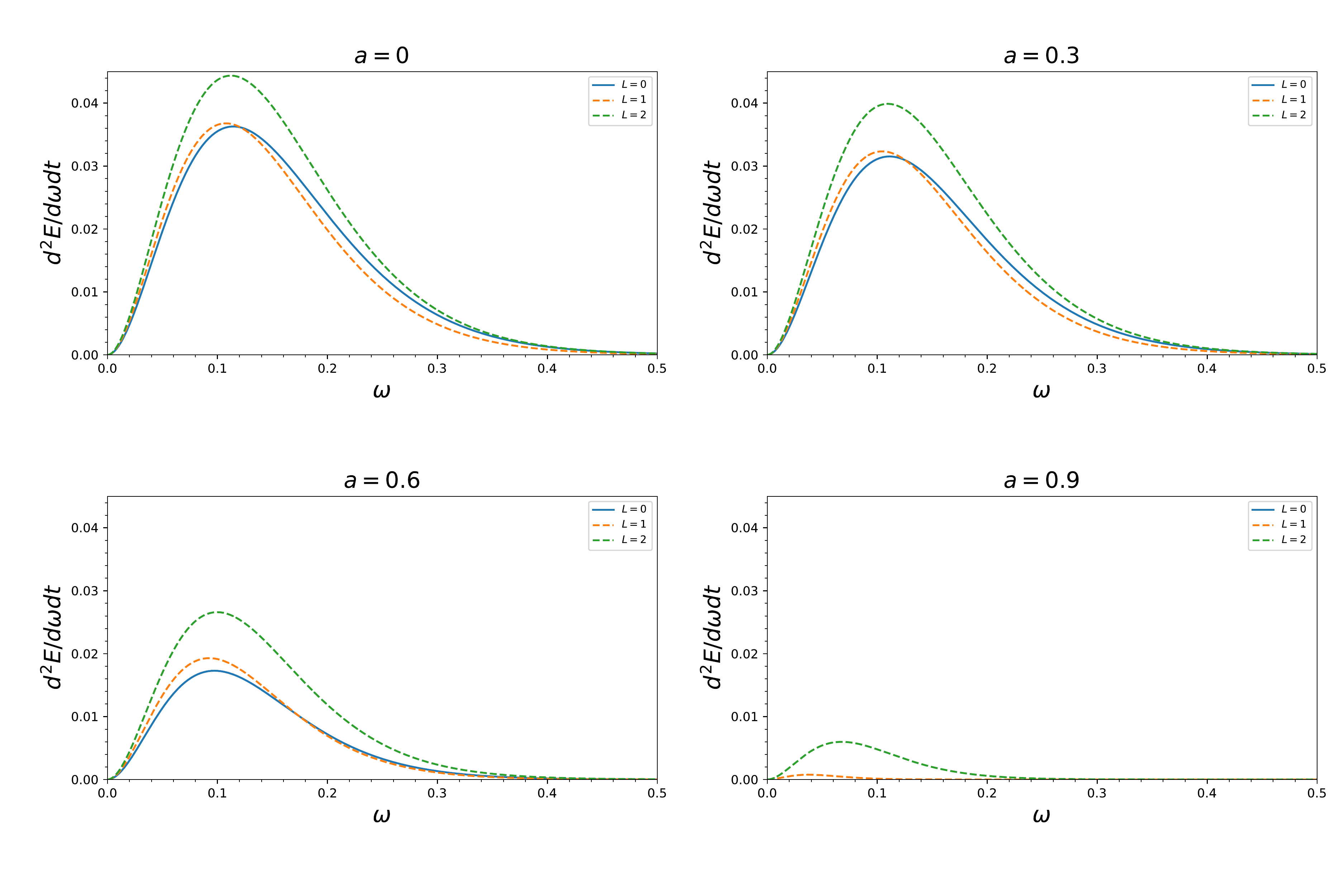}
   \caption{The energy emissivity changes with the particle frequency under different model parameters (Model 1). Curves of various colors correspond to different $L$ values, where the deformation parameter $\alpha_{0}=0.8$.}
  \label{emission_rate_type1_Kerr_alpha0_1}
\end{figure}

\begin{figure}[htbp]
  \centering
   \includegraphics[scale=0.29]{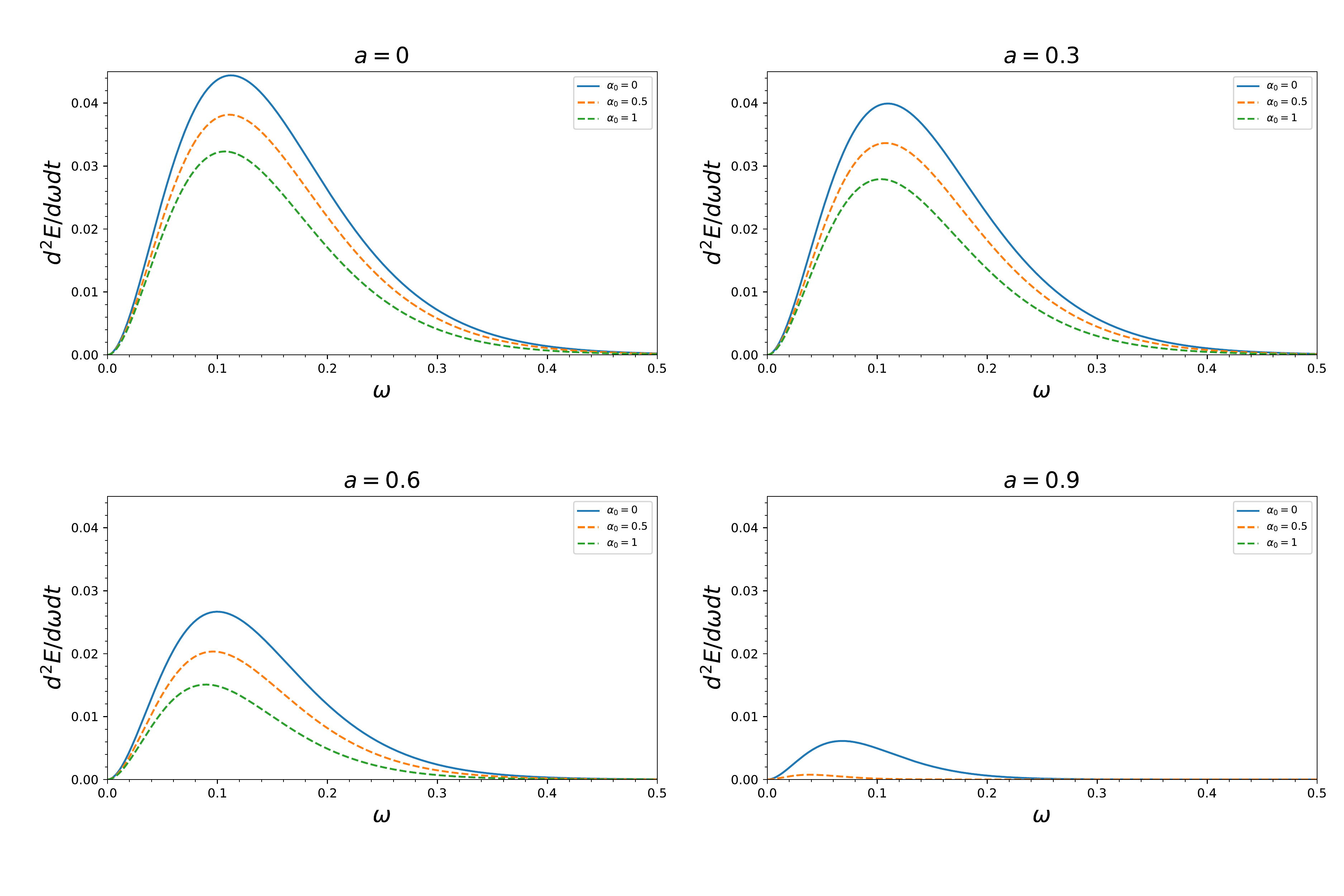}
   \caption{The energy emissivity changes with the particle frequency under different model parameters (Model 1). Curves of various colors correspond to different deformation parameters $\alpha_{0}$, where parameter $L=0.5$.}
  \label{emission_rate_type1_Kerr_L_1}
\end{figure}

\begin{figure}[htbp]
  \centering
   \includegraphics[scale=0.29]{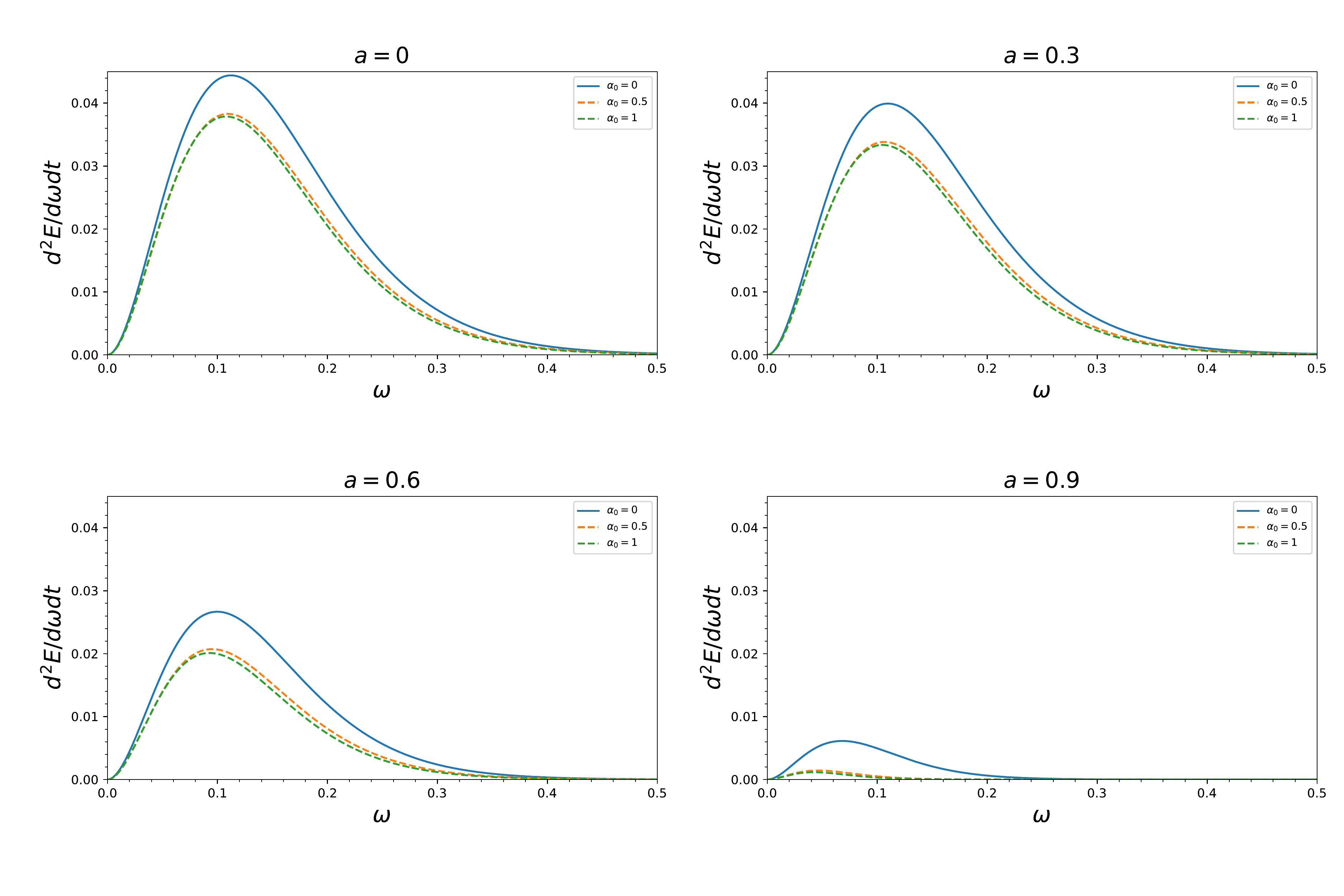}
   \caption{The energy emissivity changes with the particle frequency under different model parameters (Model 1). Curves of various colors correspond to different deformation parameters $\alpha_{0}$, where parameter $L=1$.}
  \label{emission_rate_type1_Kerr_L_2}
\end{figure}

\begin{figure}[htbp]
  \centering
   \includegraphics[scale=0.29]{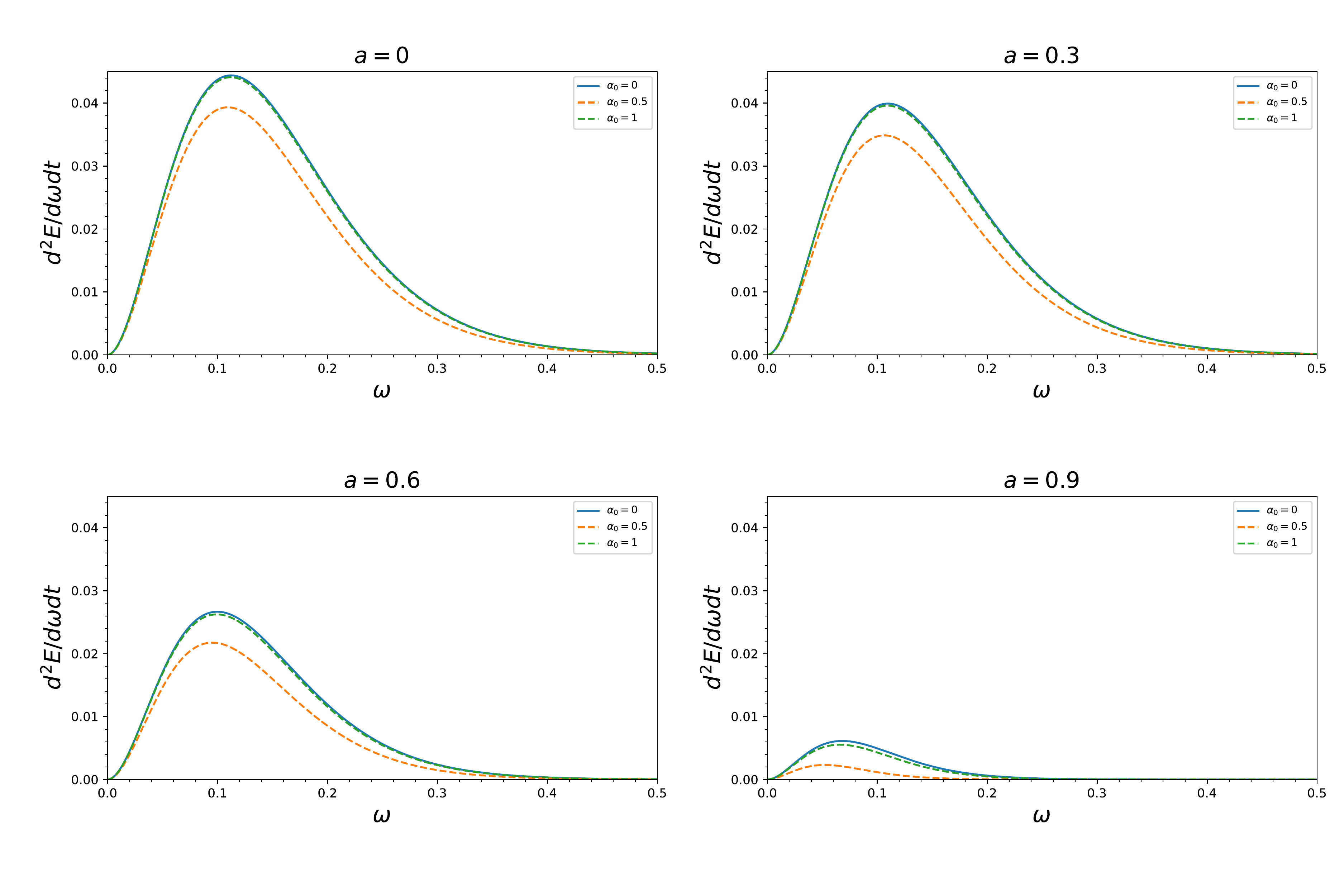}
   \caption{The energy emissivity changes with the particle frequency under different model parameters (Model 1). Curves of various colors correspond to different deformation parameters $\alpha_{0}$, where parameter $L=1.5$.}
  \label{emission_rate_type1_Kerr_L_3}
\end{figure}

\begin{figure}[htbp]
  \centering
   \includegraphics[scale=0.29]{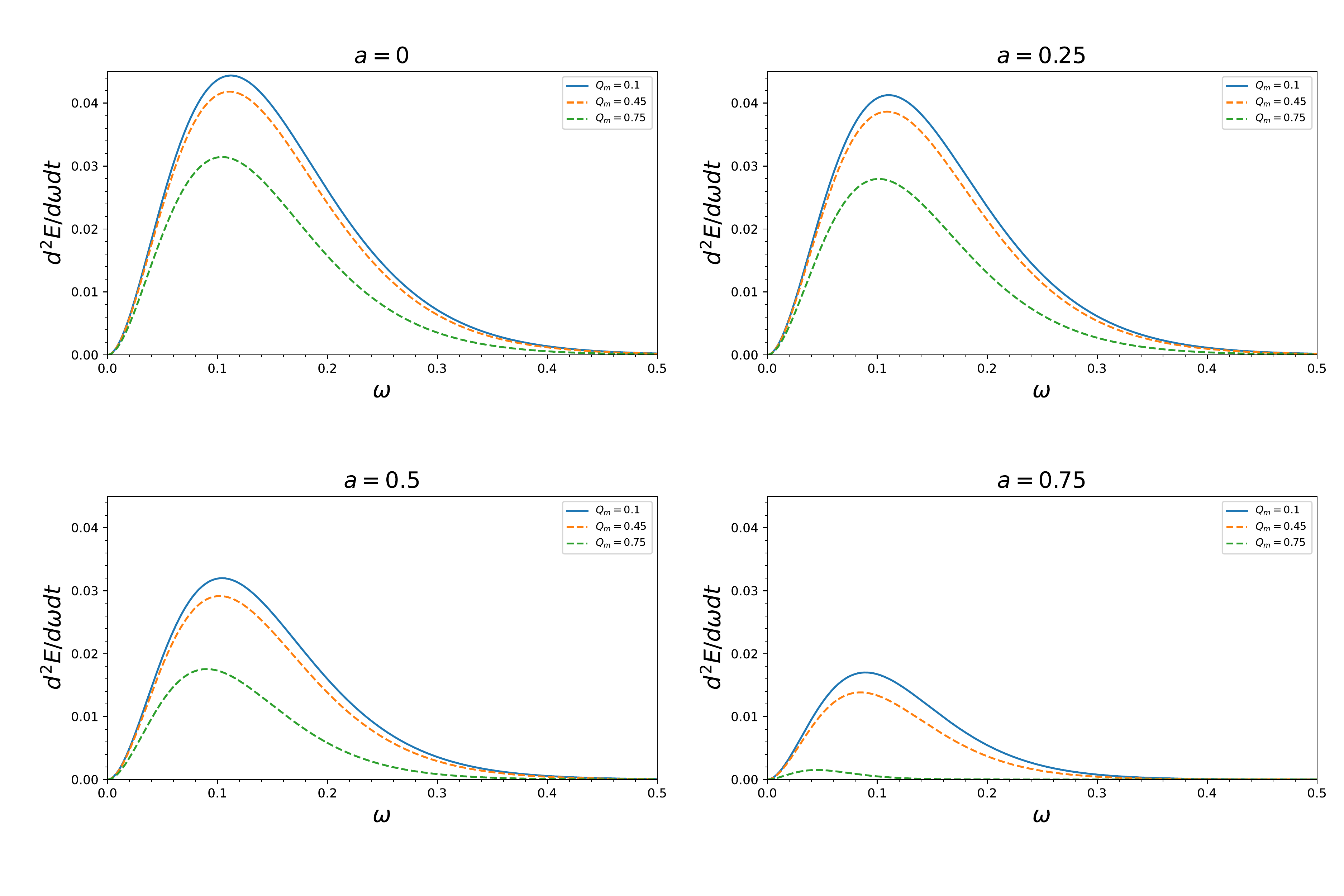}
   \caption{The energy emissivity changes with the particle frequency under different model parameters (Model 2). Curves of various colors correspond to different values of intensity parameters $Q_{m}$. Here $k=1.5$.}
  \label{emiss_rate_type2_k_1}
\end{figure}

Firstly, it can be seen from the energy emissivity formula (\ref{52}) that its form is very close to Planck black-body radiation formula, so its main numerical results (see Figures \ref{emission_rate_type1_Kerr_alpha0_1}, \ref{emission_rate_type1_Kerr_L_1}, \ref{emission_rate_type1_Kerr_L_2}, \ref{emission_rate_type1_Kerr_L_3}, \ref{emiss_rate_type2_k_1}) are close to the black-body radiation curve. However, there are some differences, $\sigma_{eim}$ is different at various Hawking temperatures. 

Secondly, in model $1$, for the case of $\alpha_{0}$ with a constant value ($\alpha_{0}=0.8$), with the increase of parameter $L$, the overall trend of its energy emissivity increases, however, a more complicated situation appears on the right side of the emissivity peak, that is, the emissivity curves corresponding to different $L$ values intersect. The intersection of energy emissivity curves may reveal a new physical process. We take Figure \ref{emission_rate_type1_Kerr_alpha0_1} as an example for analysis. When $\alpha_{0}=0.8$, the existence of the intersections indicates that the change of the energy emissivity with parameter $L$ is not monotonic. As the spin of the black hole increases, the position of the intersection shifts to the right (i.e. the greater the particle frequency $\omega$). In Figure \ref{emission_rate_type1_Kerr_L_2}, when parameter $L=1$, the energy emissivity curves corresponding to different $\alpha_{0}$ also show the phenomenon of intersections. We find that with the increase of the spin of the black hole, the position of the intersection moves to the left (that is, the direction in which the particle frequency decreases). The intersection phenomenon of energy emissivity curves is the characteristic of model $1$. For model $2$, namely the case of the black hole with short hairs, Figure \ref{emiss_rate_type2_k_1} shows the results of numerical calculation, including: (1) the energy emissivity decreases with the increase of short-hair parameter $Q_{m}$, different from model 1, the energy emissivity curves corresponding to these different $Q_{m}$ values do not exist intersection phenomenon; (2) due to the close relationship between short-hair parameter $Q_{m}$ and the black hole spin $a$, the value of $Q_{m}$ is greatly limited.  
 
Finally, comparing model $1$ and model $2$, the main difference of energy emissivity is whether there is intersection phenomenon, which provides a new possibility for EHT to measure the black hole hairs. That is, whether a black hole has a normal or short hair should show up in EHT measurements.

\section{Summary}
\label{6}

In this work, we obtain the exact solution (space-time linear element (2.25)) of the short-hair black hole in the rotation case. Combined with the space-time metric of the black hole with hairs, we calculate the influence of short hair and  hairs on the black hole shadows in detail, and analyse the physical significance of these results. Specifically, our main results are as follows:

(1) Using the NJ method, we generalize the spherically symmetric short-hair black hole metric to the rotation case (space-time linear element (2.25)). In the case of spherically symmetric short-hair black hole, the value range of parameter $k$ is $k>1$. For the rotational short-hair black hole (2.25), the range of short-hair charge value $Q_{m}$ is greatly reduced due to the introduction of the black hole spin $a$. 
When $0\leqslant Q_{m}\leqslant \frac{2}{3}\times 4^{\frac{1}{3}}$, the rotational short-hair black hole has three event horizons, two of which are equal, so from physical observation, the black hole has only two event horizons at this time. When $Q_{m}> \frac{2}{3}\times 4^{\frac{1}{3}}$, the rotational short-hair black hole has three unequal event horizons, so the space-time structure of the black hole is significantly different from that of Kerr black hole.
In the space-time metric (2.25), the value range of $k$ is $k>1$. If the value of $k$ is larger, some interesting phenomena will occur in the rotational short-hair black hole. We take $k=\frac{5}{2}$ as an example for rough analysis. When $k=\frac{5}{2}$, the roots of equations (\ref{13}) and (2.25) will be very complex; As long as the values of $Q_{m}$ and $a$ are appropriate, short-hair black holes may have more event horizons. Then, what is the special physical meaning of the short-hair black holes with multiple event horizons (especially more than $3$)? We look forward to studying this issue in future work. 

(2) We calculate the shadow shapes for two kinds of black holes with hairs in detail, and further study the scale, distortion properties and energy emissivity of black hole shadows.
First of all, for model $1$, the effect of scalar hair on the black hole shadows corresponds to that of $\varepsilon>0$ in reference\citep{2021JCAP...09..028K}, but the specific changes of the shadows in model $1$ are different. This is because the black hole hair in reference\citep{2021JCAP...09..028K} is considered as a perturbation to the black hole, while the space-time metric of model $1$ is accurate and does not have perturbation property. For model $2$, that is, the change of the black hole shadow caused by short hairs, the main change trend is consistent with that of $\varepsilon<0$ in reference\citep{2021JCAP...09..028K}. Because of the special structure of the short-hair black hole, the specific changes of black hole shadows are different.
Secondly, the variation of $R_{s}$ and $\delta_{s}$ with $L$ and $\alpha_{0}$ is not a monotone function in model $1$, but in model $2$, it is. These results show that scalar hairs (model $1$) have different effects on Kerr black hole shadows than short hairs (model $2$), so it is possible to distinguish the types and properties of these hairs if they are detected by EHT observations. 
Finally, as for the effects of the hairs on energy emissivity, the main results in model $1$, different energy emissivity curves have intersection phenomenon, while in model $2$ (short-hair black hole), there is no similar intersection phenomenon.      

In general, various black hole hairs have different effects on the shadows, such as non-monotonic properties and intersection phenomena mentioned in this paper. Using these characteristics, it is possible to test the no-hair theorem in future EHT observations, so as to have a deeper understanding of the quantum effect of black holes. In future work, we will use numerical simulations to study the effects of various hairs on black holes and their observed properties.

\begin{acknowledgments}
We acknowledge the anonymous referee for a constructive report that has significantly improved this paper. We acknowledge the  Special Natural Science Fund of Guizhou University (grant
No. X2020068) and the financial support from the China Postdoctoral Science Foundation funded project under grants No. 2019M650846.
\end{acknowledgments}


\nocite{*}


\begin{thebibliography}{}


\bibitem{1973grav.book.....M}
Misner, C.~W., Thorne, K.~S., \& Wheeler, J.~A., \emph{Gravitation}, \emph{San Francisco: W.H. Freeman and Co.} (1973)
\bibitem{1998bhp..book.....F}
Frolov, V.~P. \& Novikov, I.~D., \emph{Black hole physics: basic concepts and new developments}, \emph{https://ui.adsabs.harvard.edu/abs/1998bhp..book.....F}

\bibitem{1963PhRvL..11..237K}
Kerr, R.~P., \emph{Gravitational Field of a Spinning Mass as an Example of Algebraically Special Metrics}, \emph{PhysRevLett} {\bf 11} (1963) 237-238.

\bibitem{1984ucp..book.....W}
Wald, R., \emph{General relativity}, \emph{Chicago, University of Chicago Press} (1984) 504 p.
\bibitem{1998mtbh.book.....C}
Chandrasekhar, S., \emph{The Mathematical Theory of Black Holes}, \emph{Oxford University Press} (1998)

\bibitem{1939PhRv...56..455O}
Oppenheimer, J.~R. \& Snyder, H., \emph{On Continued Gravitational Contraction}, \emph{Physical Review} {\bf 56} (1939) 455459. 

\bibitem{1977lss..conf.....H}
Hawking, S.~W. \& Ellis, G.~F.~R., \emph{The large-scale structure of space-time}, \emph{Lunar Sample Studies} (1977)

\bibitem{1967PhRv..164.1776I}
Israel, W., \emph{Event Horizons in Static Vacuum Space-Times}, \emph{Physical Review} {\bf 164} (1967) 1776-1779. 
\bibitem{1968CMaPh...8..245I}
Israel, W., \emph{Event horizons in static electrovac space-times}, \emph{Communications in Mathematical Physics} {\bf 8} (1968) 245.
\bibitem{1971PhRvL..26..331C}
Carter, B., \emph{Axisymmetric Black Hole Has Only Two Degrees of Freedom}, \emph{PhysRevLett} {\bf 26} (1971) 331-333.
\bibitem{1972CMaPh..25..152H}
Hawking, S.~W., \emph{Black holes in general relativity}, \emph{Communications in Mathematical Physics} {\bf 25} (1972) 152-166.
\bibitem{1975PhRvL..34..905R}
Robinson, D.~C., \emph{Uniqueness of the Kerr Black Hole}, \emph{PhysRevLett} {\bf 34} (1975) 905-906.
\bibitem{1982JPhA...15.3173M}
Mazur, P.~O., \emph{Proof of uniqueness of the Kerr-Newman black hole solution}, \emph{Journal of Physics A Mathematical General} {\bf 15} (1982) 3173-3180.
\bibitem{1995PhRvD..51.6608B}
Bekenstein, J.~D., \emph{Novel ``no-scalar-hair'' theorem for black holes}, \emph{PhysRevD} {\bf 51} (1995) R6608-R6611.
\bibitem{2014ApJ...784....7B}
Broderick, A.~E., Johannsen, T., Loeb, A., et al., \emph{Testing the No-hair Theorem with Event Horizon Telescope Observations of Sagittarius A*}, \emph{Astrophys.J} {\bf 784} (2014) 7. arXiv:1311.5564.
\bibitem{2019PhRvL.123k1102I}
Isi, M., Giesler, M., Farr, W.~M., et al., \emph{Testing the No-Hair Theorem with GW150914}, \emph{PhysRevLett} {\bf 123} (2019) 111102. arXiv:1905.00869.
\bibitem{2021arXiv211100953W}
Wang, K., \emph{Retesting the no-hair theorem with GW150914},  (2021). arXiv:2111.00953.
\bibitem{2015PhRvL.114o1102G}
G{\"u}rlebeck, N., \emph{No-Hair Theorem for Black Holes in Astrophysical Environments}, \emph{PhysRevLett} {\bf 114} (2015) 151102. arXiv:1503.03240.
\bibitem{2015IJMPD..2442014H}
Herdeiro, C.~A.~R. \& Radu, E., \emph{Asymptotically flat black holes with scalar hair: A review}, \emph{International Journal of Modern Physics D} {\bf 24} (2015) 1542014-219. arXiv:1504.08209.

\bibitem{2016PhRvL.116w1301H}
Hawking, S.~W., Perry, M.~J., \& Strominger, A., \emph{Soft Hair on Black Holes}, \emph{PhysRevLett} {\bf 116} (2016) 231301. arXiv:1601.00921.
\bibitem{2014PhRvL.112v1101H}
Herdeiro, C.~A.~R. \& Radu, E., \emph{Kerr Black Holes with Scalar Hair}, \emph{PhysRevLett} {\bf 112} (2014) 221101. arXiv:1403.2757.
\bibitem{2015CQGra..32n4001H}
Herdeiro, C. \& Radu, E., \emph{Construction and physical properties of Kerr black holes with scalar hair}, \emph{Classical and Quantum Gravity} {\bf 32} (2015) 144001. arXiv:1501.04319.
\bibitem{2013PhLB..719..419D}
Dvali, G. \& Gomez, C., \emph{Black hole's 1/N hair}, \emph{Physics Letters B} {\bf 719} (2013) 419-423. arXiv:1203.6575.
\bibitem{1992NuPhB.378..175C}
Coleman, S., Preskill, J., \& Wilczek, F., \emph{Quantum hair on black holes}, \emph{Nuclear Physics B} {\bf 378} (1992) 175-246. arXiv:hep-th/9201059.
\bibitem{1991PhST...36..258P}
Preskill, J., \emph{Quantum hair}, \emph{Physica Scripta Volume T} {\bf 36} (1991) 258-264.
\bibitem{2017CQGra..34t4001B}
Bousso, R. \& Porrati, M., \emph{Soft hair as a soft wig}, \emph{Classical and Quantum Gravity} {\bf 34} (2017) 204001. arXiv:1706.00436.

\bibitem{2021PDU....3100744O}
Ovalle, J., Casadio, R., Contreras, E., et al., \emph{Hairy black holes by gravitational decoupling}, \emph{Physics of the Dark Universe} {\bf 31} (2021) 100744. arXiv:2006.06735.
\bibitem{2021PhRvD.103d4020C}
Contreras, E., Ovalle, J., \& Casadio, R., \emph{Gravitational decoupling for axially symmetric systems and rotating black holes}, \emph{PhysRevD} {\bf 103} (2021) 044020. arXiv:2101.08569.

\bibitem{2011PhRvD..83l4015J}
Johannsen, T. \& Psaltis, D., \emph{Metric for rapidly spinning black holes suitable for strong-field tests of the no-hair theorem}, \emph{PhysRevD} {\bf 83} (2011) 124015. arXiv:1105.3191.
\bibitem{2020JCAP...09..026K}
Khodadi, M., Allahyari, A., Vagnozzi, S., et al., \emph{Black holes with scalar hair in light of the Event Horizon Telescope}, \emph{Journal of Cosmology and Astroparticle Physics} {\bf 2020} (2020) 026. arXiv:2005.05992.
\bibitem{1997IJMPD...6..563D}
David, J., Husain, V., \emph{Black Holes with Short Hair}, \emph{International Journal of Modern Physics D} {\bf 6} (1997) 563-573. arXiv:gr-qc/9707027.

\bibitem{2016PhRvL.116f1102A}
Abbott, B.~P., Abbott, R., Abbott, T.~D., et al., \emph{Observation of Gravitational Waves from a Binary Black Hole Merger}, \emph{PhysRevLett} {\bf 116} (2016) 061102. arXiv:1602.03837.
\bibitem{2019ApJ...875L...1E}
Event Horizon Telescope Collaboration, Akiyama, K., Alberdi, A., et al., \emph{First M87 Event Horizon Telescope Results. I. The Shadow of the Supermassive Black Hole}, \emph{Astrophysical Journal Letters} {\bf 1} (2019) L1. arXiv:1906.11238.
\bibitem{2019ApJ...875L...5E}
Event Horizon Telescope Collaboration, Akiyama, K., Alberdi, A., et al., \emph{First M87 Event Horizon Telescope Results. V. Physical Origin of the Asymmetric Ring}, \emph{Astrophysical Journal Letters} {\bf 875} (2019) L5. arXiv:1906.11242.
\bibitem{2021PhRvD.103l2002A}
Abbott, R., Abbott, T.~D., Abraham, S., et al., \emph{Tests of general relativity with binary black holes from the second LIGO-Virgo gravitational-wave transient catalog}, \emph{PhysRevD} {\bf 12} (2021) 122002. arXiv:2010.14529.
\bibitem{2020PhRvL.125n1104P}
Psaltis, D., Medeiros, L., Christian, P., et al., \emph{Gravitational Test beyond the First Post-Newtonian Order with the Shadow of the M87 Black Hole}, \emph{PhysRevLett} {\bf 125} (2020) 141104. arXiv:2010.01055.
\bibitem{2021ApJ...910L..13E}
Event Horizon Telescope Collaboration, Akiyama, K., Algaba, J.~C., et al., \emph{First M87 Event Horizon Telescope Results. VIII. Magnetic Field Structure near The Event Horizon}, \emph{Astrophysical Journal Letters} {\bf 910} (2021) L13. arXiv:2105.01173.

\bibitem{2021JCAP...09..028K}
Khodadi, M., Lambiase, G., \& Mota, D.~F., \emph{No-hair theorem in the wake of Event Horizon Telescope}, \emph{Journal of Cosmology and Astroparticle Physics} {\bf 2021} (2021) 028. arXiv:2107.00834.
\bibitem{2015PhRvL.115u1102C}
P.~V.~P., Herdeiro, C.~A.~R., Radu, E., et al., \emph{Shadows of Kerr Black Holes with Scalar Hair}, \emph{PhysRevLett} {\bf 115} (2015) 211102. arXiv:1509.00021.

\bibitem{Newman:1965tw}
Newman, E. T. and Janis, A. I., \emph{Note on the Kerr spinning particle metric}, \emph{J. Math. Phys}{\bf 6} (1965) 915-917. 
\bibitem{2014PhRvD..90f4041A}
Azreg-A{\"\i}nou, M., \emph{Generating rotating regular black hole solutions without complexification}, \emph{Phys.Rev. D}{\bf 90} (2014) 064041. arXiv:1405.2569.
\bibitem{2014PhLB..730...95A}
Azreg-A{\"\i}nou, M., \emph{Regular and conformal regular cores for static and rotating solutions}, \emph{Physics Letters B}{\bf 730} (2014) 95-98. arXiv:1401.0787.

\bibitem{1968PhRv..174.1559C}
Carter, Brandon., \emph{Global Structure of the Kerr Family of Gravitational Fields}, \emph{Physical Review} {\bf 174} (1968) 1559-1571. 
\bibitem{Chandrasekhar:1985kt}
Chandrasekhar, Subrahmanyan., \emph{The mathematical theory of black holes}, \emph{https://inspirehep.net/literature/224457}, (1985).

\bibitem{2009PhRvD..80b4042H}
Hioki, Kenta \& Maeda, Kei-Ichi., \emph{Measurement of the Kerr spin parameter by observation of a compact object's shadow}, \emph{PhysRevD} {\bf 80} (2009) 024042. arXiv:0904.3575.
\bibitem{1973PhRvD...7.2807M}
Mashhoon, B., \emph{Scattering of Electromagnetic Radiation from a Black Hole}, \emph{PhysRevD} {\bf 7} (1973) 2807-2814. arXiv:0904.3575.
\bibitem{2013JCAP...11..063W}
Wei, S.-W. \& Liu, Y.-X., \emph{Observing the shadow of Einstein-Maxwell-Dilaton-Axion black hole}, \emph{Journal of Cosmology and Astroparticle Physics} {\bf 2013} (2013) 063. arXiv:1311.4251.


















%
%
%
%
%
%
%
%
%
%
%
%
%
%
%
%
%
%
%
%
%
%
%
%
%
%
%
%
%
%
%
%
%
%
%
%
%
%
%
%
%
%
%
%
%
%
%
%
%
%
%
%
%
%
%
%
%
%
%
%
%
%
%
%
%
%
%
%
%
%
%
%
%
%
%
%
%
%
%
%
%
%
%
%
%
%
%
%
%
%
%
%
%
%
%
%
%
%
%
%
%
%
%
%
%
%
%
%
%
%
%
%
%
%
%
%
%
%
%
%
%
%

\end{thebibliography}
\end{document}